\PassOptionsToPackage{unicode}{hyperref}
\PassOptionsToPackage{hyphens}{url}
\PassOptionsToPackage{dvipsnames,svgnames,x11names}{xcolor}
\documentclass[
  10pt,
]{article}
\usepackage{fontspec} 
\usepackage{xcolor}
\usepackage{amsmath,amssymb}
\usepackage{iftex}
\ifPDFTeX
  \usepackage[T1]{fontenc}
  \usepackage[utf8]{inputenc}
  \usepackage{textcomp} 
\else 
  \usepackage{unicode-math} 
  \defaultfontfeatures{Scale=MatchLowercase}
  \defaultfontfeatures[\rmfamily]{Ligatures=TeX,Scale=1}
\fi
\usepackage{lmodern}
\ifPDFTeX\else
\fi
\IfFileExists{upquote.sty}{\usepackage{upquote}}{}
\IfFileExists{microtype.sty}{
  \usepackage[]{microtype}
  \UseMicrotypeSet[protrusion]{basicmath} 
}{}
\makeatletter
\@ifundefined{KOMAClassName}{
  \IfFileExists{parskip.sty}{%
    \usepackage{parskip}
  }{
    \setlength{\parindent}{0pt}
    \setlength{\parskip}{6pt plus 2pt minus 1pt}}
}{
  \KOMAoptions{parskip=half}}
\makeatother
\usepackage{longtable,booktabs,array}
\usepackage{calc} 
\usepackage{etoolbox}
\makeatletter
\patchcmd\longtable{\par}{\if@noskipsec\mbox{}\fi\par}{}{}
\makeatother
\IfFileExists{footnotehyper.sty}{\usepackage{footnotehyper}}{\usepackage{footnote}}
\makesavenoteenv{longtable}
\usepackage{graphicx}
\makeatletter
\newsavebox\pandoc@box
\newcommand*\pandocbounded[1]{
  \sbox\pandoc@box{#1}%
  \Gscale@div\@tempa{\textheight}{\dimexpr\ht\pandoc@box+\dp\pandoc@box\relax}%
  \Gscale@div\@tempb{\linewidth}{\wd\pandoc@box}%
  \ifdim\@tempb\p@<\@tempa\p@\let\@tempa\@tempb\fi
  \ifdim\@tempa\p@<\p@\scalebox{\@tempa}{\usebox\pandoc@box}%
  \else\usebox{\pandoc@box}%
  \fi%
}
\def\fps@figure{htbp}
\makeatother
\NewDocumentCommand\citeproctext{}{}

\makeatletter
 \let\@cite@ofmt\@firstofone
 \def\@biblabel#1{}
 \def\@cite#1#2{{#1\if@tempswa , #2\fi}}
\makeatother
\newlength{\cslhangindent}
\newlength{\csllabelwidth}
\newenvironment{CSLReferences}[2] 
 {\begin{list}{}{%
  \setlength{\itemindent}{0pt}
  \setlength{\leftmargin}{0pt}
  \setlength{\parsep}{0pt}
  \ifodd #1
   \setlength{\leftmargin}{\cslhangindent}
   \setlength{\itemindent}{-1\cslhangindent}
  \fi
  \setlength{\itemsep}{#2\baselineskip}}}
 {\end{list}}
\usepackage{calc}

\usepackage[letterpaper,top=1in,bottom=1in,left=1.25in,right=1.25in]{geometry}
\usepackage{microtype}
\AtBeginDocument{\hypersetup{
  pdftitle={Which reported inputs govern molecular docking reproducibility? A benchmark from reporting audit to independent re-execution},
  pdfauthor={Giap Duc Ha}
}}
\usepackage{booktabs}
\usepackage{array}
\usepackage{graphicx}
\usepackage{placeins}
\setkeys{Gin}{width=0.82\linewidth}
\usepackage{caption}
\usepackage{xcolor}
\usepackage{titlesec}
\titleformat{\section}{\large\bfseries}{\thesection}{0.7em}{}
\titlespacing*{\section}{0pt}{1.9ex plus .3ex}{0.8ex}
\titleformat{\subsection}{\normalsize\bfseries}{\thesubsection}{0.6em}{}
\titlespacing*{\subsection}{0pt}{1.2ex plus .2ex}{0.4ex}
\titleformat{\subsubsection}{\normalsize\bfseries\itshape}{\thesubsubsection}{0.5em}{}
\titlespacing*{\subsubsection}{0pt}{0.9ex}{0.3ex}
\usepackage{fancyhdr}
\fancypagestyle{plain}{\fancyhf{}\fancyfoot[C]{\normalsize\thepage}}
\fancypagestyle{firstpage}{\fancyhf{}\fancyfoot[C]{\normalsize\thepage}\fancyfoot[L]{\footnotesize Preprint.}}
\usepackage{enumitem}\setlist{itemsep=1pt,topsep=3pt,leftmargin=1.5em}

\usepackage{bookmark}
\IfFileExists{xurl.sty}{\usepackage{xurl}}{} 
\hypersetup{
  colorlinks=true,
  linkcolor={blue},
  filecolor={Maroon},
  citecolor={blue},
  urlcolor={blue},
  pdfcreator={LaTeX via pandoc}}

\author{}
\date{}

\begin{document}
\thispagestyle{firstpage}
\begin{center}
{\noindent\rule{\linewidth}{1.6pt}}\\[3pt]
{\Large\bfseries Which reported inputs govern\\ molecular docking reproducibility?\par}
\vspace{4pt}
{\normalsize\itshape A benchmark from reporting audit to independent re-execution\par}
\vspace{6pt}
{\noindent\rule{\linewidth}{0.5pt}}\\[13pt]
{\large\bfseries Giap Duc Ha\par}
\vspace{4pt}
{New Science Lab\par}
{Nanjing Medical University\par}
\end{center}
\vspace{15pt}

\begin{center}{\large\bfseries Abstract}\end{center}

\begingroup\leftskip=2.4em \rightskip=2.4em \small

Computational docking results can be re-executed only when the molecular
system and protocol are specified, yet reporting checklists do not
quantify how strongly individual method fields affect the reported
score. We combined a span-verified audit of 50 open-access papers with
controlled one-factor perturbations, a Vinardo scoring-function check
and a Vina cross-docking extension covering 12 targets in 7 protein
families, and re-execution of 37 published claims on local and
independent cloud infrastructure. Search-related fields were sparsely
reported, but perturbing box centre, box size, exhaustiveness or random
seed produced median absolute score changes of no more than 0.08 kcal
mol\(^{-1}\); ligand protonation and identity produced median changes of
0.19 and 0.29 kcal mol\(^{-1}\), respectively, whereas receptor
structure produced the largest change (1.02 kcal mol\(^{-1}\); n = 85;
\(P < 0.001\)), and 22\% of 116 unique reported ligand strings remained
unresolved or ambiguous after deterministic name normalization. These
measurements yielded empirical field weights for the NewScience Evidence
score and identified exact receptor structure, machine-resolvable ligand
identity and protonation state as primary reporting elements; 29 of 37
local and 27 of 37 cloud re-executions were within 2.0 kcal mol\(^{-1}\)
of the reported score, but these descriptive rates do not estimate
experimental replication or a calibrated probability of reproduction.


\par\endgroup
\vspace{6pt}

\section{Introduction and Related
Work}\label{introduction-and-related-work}

A molecular docking score is the output of a specified computational
procedure. It depends on the receptor structure, ligand identity and
protonation state, search-space definition, scoring function and
stochastic search settings. Docking is widely used in computer-aided
drug discovery (Guedes et al. 2014), where reported docking scores
triage compounds for synthesis and guide claims of activity, so the
ability to re-obtain a published score from a paper's own description is
a basic form of computational reproducibility. Yet reporting of the
inputs that determine the score remains inconsistent (Jain 2008), and a
reader who wishes to re-run a published calculation is frequently unable
to recover the exact structure, the ligand or the search box that
produced it. Computational reproduction therefore requires both an
identifiable molecular system and sufficient methodological detail to
reconstruct the calculation, and it is distinct from the separate
question of whether the docking method is accurate against measured
affinities.

Reporting guidance treats docking as conditional modelling whose
interpretation depends on structural provenance, ligand-state definition
and search-space design (Kittelson et al. 2026). These recommendations
specify which fields should be reported, but they do not quantify how
strongly each field affects a re-executed score. This distinction
matters when reporting criteria are converted into a weighted
reproducibility score. We therefore measured reporting completeness,
quantified the effect of controlled perturbations to individual inputs,
tested the resulting hierarchy across scoring functions and targets, and
used the measured effects to derive empirical field weights for NSE
(Fig. 1).

\begin{figure}
\centering
\includegraphics[width=1\linewidth,height=\textheight,keepaspectratio,alt={Molecular benchmark definition and analysis design. (a) Representative carbonic anhydrase II-FB2 complex (PDB 4YX4) showing the exact receptor conformer, the machine-resolvable ligand identity and the reported search volume (translucent cube) used in the benchmark; the histidine triad coordinating the catalytic zinc (dashed bonds) is drawn as sticks. The structural vignette illustrates recorded inputs and is not an additional docking result (Gaspari et al. 2016). (b) The perturbed input classes comprised receptor conformer, ligand identity or state and search specification; the receptor overlay shows the 6 carbonic-anhydrase structures used in the original cross-docking ensemble. (c) The three analysis arms comprised a span-verified reporting audit, controlled one-factor perturbations and independent re-execution; the inset previews the measured input-effect hierarchy (median \textbar\textbackslash Delta S\textbar: receptor 1.02, ligand identity 0.29, protonation 0.19, search box \textbackslash leq 0.08 kcal mol\^{}\{-1\}; Fig. 2).}]{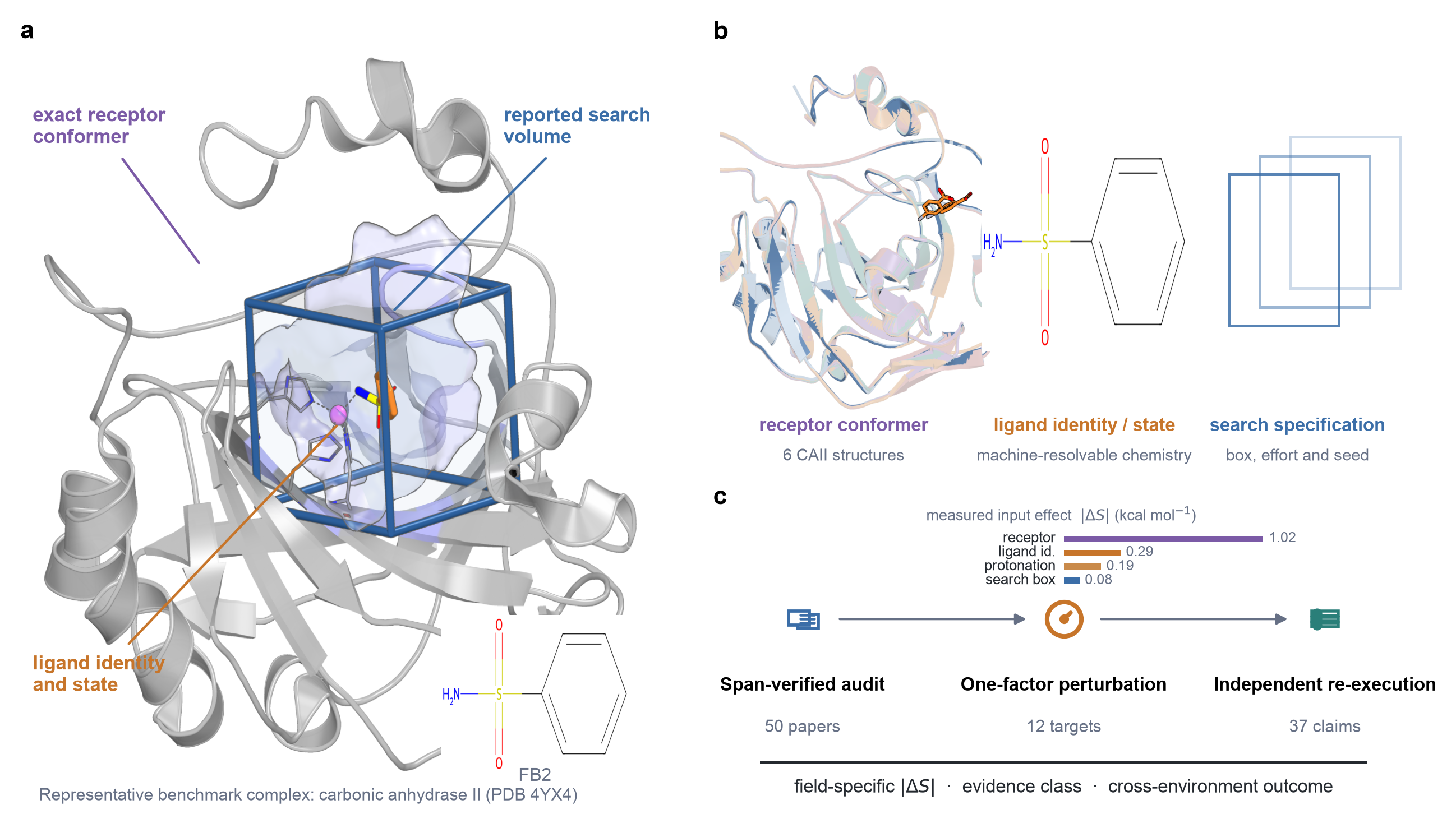}
\caption{Molecular benchmark definition and analysis design. (a)
Representative carbonic anhydrase II-FB2 complex (PDB 4YX4) showing the
exact receptor conformer, the machine-resolvable ligand identity and the
reported search volume (translucent cube) used in the benchmark; the
histidine triad coordinating the catalytic zinc (dashed bonds) is drawn
as sticks. The structural vignette illustrates recorded inputs and is
not an additional docking result (Gaspari et al. 2016). (b) The
perturbed input classes comprised receptor conformer, ligand identity or
state and search specification; the receptor overlay shows the 6
carbonic-anhydrase structures used in the original cross-docking
ensemble. (c) The three analysis arms comprised a span-verified
reporting audit, controlled one-factor perturbations and independent
re-execution; the inset previews the measured input-effect hierarchy
(median \(|\Delta S|\): receptor 1.02, ligand identity 0.29, protonation
0.19, search box \(\leq 0.08\) kcal mol\(^{-1}\); Fig. 2).}
\end{figure}

\subsection{Related work}\label{related-work}

\textbf{Reporting and minimum-information standards.} Structured
per-field disclosure has precedent across quantitative biology, from
MIAME for microarray experiments ({Brazma et al.} 2001) to ARRIVE for
animal research ({Percie du Sert et al.} 2020), and is generalised by
the FAIR principles for findable, accessible, interoperable and reusable
data ({Wilkinson et al.} 2016). Machine learning uses analogous model
cards (Mitchell et al. 2019) and datasheets (Gebru et al. 2021) to
document provenance and intended use. Docking checklists extend this
approach (Kittelson et al. 2026). These standards specify which fields
should be present but treat the fields as a flat list, so a checklist
score implicitly weights every field equally. The present study compares
selected reporting fields on a common scale of score change under
specified computational perturbations.

\textbf{Docking benchmarks.} A mature line of work evaluates scoring and
sampling accuracy against curated reference data: PDBbind assembles
measured affinities for crystallographic complexes (Wang et al. 2004),
CASF-2016 grades scoring, ranking, docking, and screening power on that
basis (Su et al. 2019), and DUD-E provides matched decoys for
virtual-screening assessment (Mysinger et al. 2012). These benchmarks
ask which method docks or scores better, using engines such as AutoDock
Vina (Trott and Olson 2010) and its Vinardo re-parametrization (Quiroga
and Villarreal 2016), and empirical scoring functions calibrated on such
data (Koes et al. 2013). Their metrics (scoring, ranking, docking and
screening power) are all defined against an external reference of
measured or known-active binding. The reproducibility question is
orthogonal: given only what a paper reports, can its published score be
re-obtained? A method can rank actives well on CASF yet remain
impossible to reproduce when the paper omits the receptor structure or
names the ligand ambiguously. Re-executability must therefore be
measured from the reporting record rather than inferred from benchmark
accuracy.

\textbf{Redocking and cross-docking reproducibility.} Critical
assessments of docking programs and scoring functions have long
documented pose and score variability across tools and targets (Warren
et al. 2006; Wang et al. 2016), and cross-docked data sets quantify the
penalty of docking into non-cognate receptor structures (Francoeur et
al. 2020). Single determinants have been isolated, notably protonation,
tautomer and stereoisomer state (Brink and Exner 2009). We extend this
literature by perturbing one reported field at a time within a fixed
engine and comparing all fields on a common effect scale.

\textbf{The computational reproducibility crisis.} Surveys report that
most researchers have failed to reproduce published results (Baker
2016), and policy statements distinguish re-execution of reported
computation from independent replication (Peng 2011) and call for the
artifacts that make re-execution possible (Stodden et al. 2016; Sandve
et al. 2013). Even when code is shared, a large fraction of published
computational analyses cannot be re-run without manual repair (Samuel
and Mietchen 2023). Our scope is the narrower of these problems:
in-silico computational re-execution, with wet-lab replication out of
scope.

\textbf{LLM extraction.} Large language models now support structured
information extraction from scientific text (Dagdelen et al. 2024),
which we adopt under a programmatic span-verification gate so that every
extracted field carries a verbatim source span rather than a free-text
assertion.

\subsection{Study overview}\label{study-overview}

The study comprised six linked analyses.

\begin{enumerate}
\def\labelenumi{\arabic{enumi}.}
\item
  We measured reporting completeness for 12 method fields in a
  multi-topic open-access corpus using span-verified extraction and
  Wilson confidence intervals (Table 1).
\item
  We quantified the effect of changing one reported input at a time
  relative to a 2.0 kcal mol\(^{-1}\) reproduction tolerance and
  measured computational noise floors (Table 2).
\item
  We tested the ordering of input effects with a second scoring function
  and across 12 targets in 7 protein families (Table 3).
\item
  We estimated the proportion of reported ligand names that could not be
  resolved to a chemical structure.
\item
  We derived empirical NSE field weights from measured score changes and
  assigned an evidence class to each of the 16 fields (Table 4).
\item
  We described reproduction outcomes across executability strata without
  treating those strata as a predictive model (Table 5).
\end{enumerate}

\section{The NSE benchmark}\label{the-nse-benchmark}

The NewScience Evidence score (NSE) is a per-field reproducibility
instrument for molecular docking reports. It is built around four
quantities that reproducibility scores commonly conflate and that we
keep separate: reporting readiness, reproduction outcome, topical
relevance and verification priority (Table 6). Reporting readiness is a
property of the document, reproduction outcome is a property of a
re-executed calculation, relevance is defined for a paper--query pair,
and priority is a decision made under a finite verification budget.
Unlike reporting standards such as MIAME ({Brazma et al.} 2001),
datasheets (Gebru et al. 2021), model cards (Mitchell et al. 2019) and
ARRIVE ({Percie du Sert et al.} 2020), NSE estimates field weights from
controlled perturbations, and it evaluates whether a reported
computation can be reconstructed rather than ranking docking methods
against binding-affinity measurements as in CASF or PDBbind (Su et al.
2019; Wang et al. 2004).

\subsection{Four estimands and their
separation}\label{four-estimands-and-their-separation}

NSE tracks four estimands and never equates them (Table 6). Reporting
readiness \(\theta_j\in[0,1]\) measures how completely claim \(j\)
discloses its docking method; the reproduction outcome \(Y_j\in\{0,1\}\)
records whether a re-executed claim reproduces within tolerance; topical
relevance \(R_j(q)\in[0,1]\) describes whether paper \(j\) addresses
query \(q\); and the best-evidence score \(B_j(q)\) and verification
priority \(P_j\) are query-conditioned rankings built on these
quantities. A calibrated reproduction probability
\(p_j=P(Y_j=1\mid s_j)\) is defined but, as shown in Section 2.7, is not
estimated at the present sample size.

Three invariants follow. First, disclosure is not reproduction:
\(\theta\) is validated against \(Y\) through the calibration map of
Section 2.7 and is never set equal to it, because a fully reported claim
can still fail to reproduce and a sparsely reported claim can succeed.
Second, a calibrated probability is a fitted object that requires
reproduction outcomes and is not \(\theta\) rescaled; until the
estimability conditions of Section 2.7 are met we report a descriptive
rate rather than a probability. Third, \(B\) and \(P\) are
query-conditioned and must pass the relevance gate of Section 2.5,
whereas \(\theta\) and \(Y\) are claim-intrinsic and must not be
contaminated by query terms. This separation is the contract that every
downstream expression respects.

\textbf{Table 6. The four NSE estimands kept separate throughout the
benchmark. Reporting readiness and reproduction outcome are intrinsic to
a claim; best-evidence and verification priority are conditioned on a
query. The calibrated reproduction probability is defined but is not
estimated in this work.}

{\def\LTcaptype{none} 
\begin{longtable}[]{@{}
  >{\raggedright\arraybackslash}p{(\linewidth - 6\tabcolsep) * \real{0.2500}}
  >{\raggedright\arraybackslash}p{(\linewidth - 6\tabcolsep) * \real{0.2500}}
  >{\raggedright\arraybackslash}p{(\linewidth - 6\tabcolsep) * \real{0.2500}}
  >{\raggedright\arraybackslash}p{(\linewidth - 6\tabcolsep) * \real{0.2500}}@{}}
\toprule\noalign{}
\begin{minipage}[b]{\linewidth}\raggedright
Symbol
\end{minipage} & \begin{minipage}[b]{\linewidth}\raggedright
Quantity
\end{minipage} & \begin{minipage}[b]{\linewidth}\raggedright
Property of
\end{minipage} & \begin{minipage}[b]{\linewidth}\raggedright
Source
\end{minipage} \\
\midrule\noalign{}
\endhead
\bottomrule\noalign{}
\endlastfoot
\(\theta_j\) & reporting readiness & the document & five-item subscale
of the 12-field audit (Section 2.4) \\
\(Y_j\) & reproduction outcome & a re-executed calculation &
reproduction harness (Section 3.6) \\
\(R_j(q)\) & topical relevance & a paper--query pair & retrieval /
audit \\
\(B_j(q),\ P_j\) & best-evidence, priority & a paper--query pair &
aggregation (Sections 2.5, 2.8) \\
\end{longtable}
}

\subsection{The minimum executable reporting
set}\label{the-minimum-executable-reporting-set}

NSE contains 16 fields that describe the physical inputs, ligand, search
specification, docking engine, reported outcome and reproducibility
practices (Table 4). Twelve fields were included in the reporting audit
and verified against source-text spans (Table 1). Ligand identity was
evaluated by the resolvability analysis, and protonation was evaluated
by controlled perturbation. Water and ion handling, receptor preparation
and code artifacts were not tested directly. For descriptive purposes,
reporting frequencies were classified as near-universal
(\textgreater90\%), well reported (60--90\%), sparse (20--50\%) or rare
(\textless20\%).

\subsection{The E0 to E4 executability
ladder}\label{the-e0-to-e4-executability-ladder}

Each paper is placed on a five-level executability ladder by an
operational rule over the verified reporting record. Let \(T\), \(L\),
and \(S\) indicate that the target structure, ligand identity, and
binding site are recoverable, respectively; a binding site can be
recovered from numeric box coordinates, a named co-crystallized ligand,
or named pocket residues. Let \(X\) indicate that the remaining
execution parameters are explicitly reported, and let \(V\) indicate a
reported robustness check:

\begin{equation}
E(\text{paper})=\begin{cases} E_0 & \text{no usable engine-and-result method record} \\ E_1 & \neg(T\wedge L\wedge S) \quad\text{(a critical input is unrecoverable)} \\ E_4 & T\wedge L\wedge S\wedge X\wedge V \quad\text{(directly executable plus robustness)} \\ E_3 & T\wedge L\wedge S\wedge X \quad\text{(directly executable as reported)} \\ E_2 & T\wedge L\wedge S\wedge\neg X \quad\text{(executable only under explicit assumptions).} \end{cases}
\end{equation}

A paper is assigned to \(E_1\) when at least one critical input is
unrecoverable, to \(E_2\) when all critical inputs are recoverable but
at least one execution parameter is omitted, to \(E_3\) when the
calculation is executable as reported, and to \(E_4\) when an \(E_3\)
record also reports robustness or validation. This classification
describes the reporting record and does not evaluate scientific
correctness.

\subsection{A measurement model for reporting
readiness}\label{a-measurement-model-for-reporting-readiness}

Reporting readiness is scored with a measurement model rather than with
asserted weights, because asserted weights invite the immediate
objection of why those particular numbers were chosen. Claim \(j\) is
audited on \(k\) binary reporting items \(X_{ji}\in\{0,1\}\), and
internal consistency is summarised by Cronbach's alpha,

\begin{equation}
\alpha=\frac{k}{k-1}\left(1-\frac{\sum_i \operatorname{Var}(X_i)}{\operatorname{Var}\!\left(\sum_i X_i\right)}\right).
\end{equation}

The full 12-item set gave \(\alpha=0.62\), depressed by two
uninformative items: the docking engine, reported in 50 of 50 papers
with zero variance, and the random seed, reported in 1 of 50 and
therefore near-degenerate. We retained seven fields as descriptive
checklist items and selected an exploratory five-item core subscale
(grid centre, grid size, exhaustiveness, redocking validation and
protein chain) with \(\alpha=0.73\), which is the scored scale (Cronbach
1951; Tavakol and Dennick 2011).

The exploratory item weights use corrected item--rest associations. For
comparison, under a two-parameter logistic item-response model,

\begin{equation}
P(X_i=1\mid\theta)=\sigma\!\big(a_i(\theta-b_i)\big),\qquad \sigma(z)=\frac{1}{1+e^{-z}},
\end{equation}

each item carries Fisher information

\begin{equation}
I_i(\theta)=a_i^2\,\sigma\!\big(a_i(\theta-b_i)\big)\,\big[1-\sigma\!\big(a_i(\theta-b_i)\big)\big],
\end{equation}

so an item that better separates rigorous from sloppy reporting, that is
one with larger discrimination \(a_i\), should weigh more, with optimal
weight proportional to \(a_i^2\) at the point of maximum information. We
did not fit a two-parameter logistic model to this 50-paper audit
(Appendix B). Instead, we used the corrected item--rest correlation as
an exploratory scoring heuristic,

\begin{equation}
r_i=\operatorname{corr}\!\Big(X_i,\ \textstyle\sum_{m\neq i} X_m\Big),
\end{equation}

an empirical measure of within-subscale association, rather than an
estimate of the two-parameter logistic discrimination parameter.
Normalising the item--rest correlations gives the field weights and the
readiness score,

\begin{equation}
w_i=\frac{r_i}{\sum_\ell r_\ell},\qquad \theta_j=\sum_i w_i X_{ji}\in[0,1],
\end{equation}

with
\((w_{\text{grid centre}},w_{\text{grid size}},w_{\text{exhaustiveness}},w_{\text{redocking}},w_{\text{chain}})=(0.234,0.219,0.209,0.169,0.168)\).
The ranking that these weights produce is robust to the weighting
choice: the readiness ranking under derived weights agreed with a
hand-weighted ranking at Spearman \(\rho=0.90\) and with an unweighted
item sum at \(\rho=0.87\). These empirical weights give a transparent
exploratory scoring rule; they are not the Fisher-information weights of
a fitted item-response model. Conditional on retained weights, the score
is deterministic. The weights and subscale selection nevertheless have
sampling uncertainty, which was not quantified here; the absence of a
per-paper interval does not establish error-free measurement (Appendix
B).

\subsection{Best-evidence aggregation: axioms and the relevance
gate}\label{best-evidence-aggregation-axioms-and-the-relevance-gate}

The best-evidence score combines topical relevance \(R\in[0,1]\) and
graded quality \(Q\in[0,1]\) in a proposed score \(B\in[0,100]\). We
require monotonicity in both inputs, a zero score as relevance tends to
zero, a bounded venue contribution, and separability,
\(B=100\,g(R)\,h(Q)\). We allow an explicit relevance-threshold
discontinuity and require continuity within the admitted and excluded
relevance regions. These are design conditions, not a uniqueness
theorem.

For a nontrivial quality factor, the zero-relevance condition requires
\(g(0)=0\). An additive rule \(\alpha R+\beta Q\), with \(\beta>0\),
fails that condition because it retains \(\beta Q\) when \(R=0\). A
geometric mean \(100\sqrt{RQ}\) satisfies the zero-relevance condition
and is not excluded by these conditions. A veto caused by any missing
constituent quality field arises only if \(Q\) is itself formed as a
product of those fields; it does not follow from the outer geometric
mean. We use graded, additive quality components and select the
following hard relevance gate as a transparent operational choice:

\begin{equation}
B=100\,\mathbf{1}[R\ge\tau_R]\,R\,Q.
\end{equation}

For \(\tau_R>0\), this rule is exactly zero below threshold and is
discontinuous at the threshold. A sigmoid factor
\(\sigma(\kappa(R-\tau_R))\) instead leaves a positive value at \(R=0\);
it is a leaky approximation unless explicitly normalized or truncated.
The hard gate was chosen to make the exclusion rule inspectable, rather
than derived as the only admissible aggregator. Under this rule, the six
injected below-threshold decoy records receive zero scores by
construction. The decoy test therefore checks implementation of the rule
rather than independently validating relevance or scientific quality.
Separately, retained title-based relevance scores on 3,084 papers in 11
topics gave an AUROC of 0.987 for the designated main-protease topic
versus the remaining topic groups. That result evaluates topic
separation in this corpus, not the validity of the full best-evidence
score.

\subsection{Graded quality}\label{graded-quality}

The quality term \(Q\) is a renormalised weighted mean of method rigour
(\(M\)), reporting completeness (\(P\)), artifact availability (\(A\)),
integrity (\(I\)), citation context (\(C\)) and source (\(J\)):

\begin{equation}
Q=\frac{\sum_{d\in\mathcal{A}} w_d\, D_d}{\sum_{d\in\mathcal{A}} w_d}\in[0,1],\qquad (w_M,w_P,w_A,w_I,w_C,w_J)=(0.30,0.25,0.20,0.10,0.10,0.05).
\end{equation}

Consistent with axiom A3, the venue dimension enters only through a hard
cap, \(\partial Q/\partial J\le\varepsilon\) with \(w_J\le0.05\). The
denominator includes only assessed dimensions \(\mathcal{A}\), so
unassessed items are excluded rather than assigned a value of zero,
which prevents missing information from being read as negative evidence.
The readiness subscale of Section 2.4 contributes to the rigour
dimension. Its item-rest weights are reported as an analysis result.

\subsection{Calibration from score to reproduction
probability}\label{calibration-from-score-to-reproduction-probability}

The reproduction outcome is a thresholded readout of a re-executed
calculation,

\begin{equation}
Y_j=\mathbf{1}\!\left[\ \big|a_{\text{rep},j}-a_{\text{rerun},j}\big|\le\tau_{\text{repro}}\right],\qquad \tau_{\text{repro}}=2.0\ \text{kcal mol}^{-1},
\end{equation}

and the calibration layer seeks a monotone map from a retained ranker
score \(s_j\) to a reproduction probability,

\begin{equation}
p_j=c(s_j)=P(Y_j=1\mid s_j)\in[0,1].
\end{equation}

Two constraints govern this map. First, de-circularity: the score must
be computed without using the observed reproduction outcome or
reported-rerun error. Pre-run executability can be a predictor, but
calibration among executed claims would remain conditional on selection
into the executable set. Second, estimability: the map \(c\) can be
fitted by isotonic regression or by Platt scaling,

\begin{equation}
p=\sigma(as+b),
\end{equation}

and evaluated by the Brier score together with the expected calibration
error over \(B\) probability bins,

\begin{equation}
\mathrm{Brier}=\frac{1}{n}\sum_{j=1}^{n}(p_j-Y_j)^2,\qquad \mathrm{ECE}=\sum_{b=1}^{B}\frac{n_b}{n}\,\big|\operatorname{acc}_b-\operatorname{conf}_b\big|,
\end{equation}

with a reliability diagram and paper-clustered bootstrap intervals.
Isotonic regression overfits at small \(n\), and re-executions drawn
from one paper share a target, box and preparation and are therefore
correlated, so the binding sample size is the number of paper clusters
rather than the number of claims. At the present scale of 37
re-executions from 7 papers a stable calibration map is not
identifiable, and within the executable set the continuous quality score
did not predict the absolute reproduction error (Pearson \(r=0.10\)). We
therefore report a descriptive within-tolerance rate and not a
calibrated probability. Cross-environment score variation can change a
binary within-tolerance verdict near the 2.0 kcal mol\(^{-1}\) cutoff
and must be assessed through repeated environments or sensitivity
analysis. The measured energy differences do not establish a minimum
resolution in probability units or a universal impossibility of
calibration. The proposed larger campaign in Section 3.8 would need to
quantify both sampling uncertainty and environment-dependent outcome
variation.

\subsection{Verification priority}\label{verification-priority}

The verification-priority score allocates a finite re-execution budget
and shares the non-compensatory gate of the best-evidence score, so that
a below-threshold off-topic claim receives zero priority:

\begin{equation}
P=100\,\cdot\,\mathbf{1}\!\left[R\ge\tau_R\right]\,\cdot\,R\,\cdot\,f_{\text{feas}}(E)\,\cdot\,\mathrm{CostAdj}\,\cdot\,\big(0.45\,\mathrm{Impact}+0.30\,\mathrm{Uncertainty}+0.25\,\mathrm{Risk}\big),
\end{equation}

with \(f_{\text{feas}}(E_0,\dots,E_4)=(0.10,0.25,0.50,0.90,0.95)\). The
three value weights and feasibility factors are provisional choices.
Because the factors for E0 and E1 are positive, this expression
discounts rather than excludes unrecoverable claims; a strict executable
route would require a separate critical-input gate. A possible
uncertainty proxy is the variance of a Bernoulli reproduction outcome,

\begin{equation}
\mathrm{Uncertainty}(p)=p(1-p),
\end{equation}

which peaks at \(p=0.5\). This variance is not, by itself, expected
value of information: that requires a specified decision, loss function
and possible observation. One unvalidated utility heuristic would be

\begin{equation}
P_j\ \propto\ \mathbf{1}\!\left[R_j\ge\tau_R\right]\,f_{\text{feas}}(E_j)\,\frac{\mathrm{Uncertainty}(p_j)\,\mathrm{Impact}_j\,\mathrm{Risk}_j}{\mathrm{cost}_j},
\end{equation}

which is a different proposed rule from the weighted sum, rather than
its derived special case. Both would require external validation.
Selecting a set \(S\) of claims within a budget to maximise total value,

\begin{equation}
\max_{S}\ \sum_{j\in S} V_j\quad\text{subject to}\quad \sum_{j\in S}\mathrm{cost}_j\le \mathrm{Budget},
\end{equation}

is a knapsack problem for additive values. If verification has
diminishing returns, a set-valued submodular objective would be needed;
approximation guarantees depend on the objective, constraints and
algorithm. No budgeted selection algorithm or approximation guarantee
was tested here. Because the objective form requires the calibrated
probability \(p_j\), which is not yet estimable (Section 2.7), we
present the verification-priority axis only as a proposed companion and
do not use it as a ranking instrument in this work; an interim priority
runs on the executability gate together with impact and cost.

\subsection{Empirical field weights}\label{empirical-field-weights}

We defined each empirical method-field weight as a monotone function of
its measured effect on the reproduced score:

\begin{equation}
w_{\text{field}}=\min\!\Big(\operatorname{median}\big|\Delta S_{\text{field}}\big|,\ \tau\Big),
\end{equation}

where \(\operatorname{median}|\Delta S_{\text{field}}|\) is the median
absolute score change under a one-parameter perturbation (Table 2) and
\(\tau\) is the 2.0 kcal mol\(^{-1}\) reproduction tolerance. Receptor
structure received the largest empirical weight
(\(w_{\text{pdb}}=1.02\)), followed by ligand identity (0.29 plus a
non-compensatory resolvability prerequisite) and protonation (0.19). The
corresponding values for grid centre, grid size, random seed and
exhaustiveness were 0.08, 0.05, 0.04 and 0.01, respectively (Table 4).
Seven fields were supported by direct perturbations, 2 by partial
perturbations, 3 by reproduction outcomes and 1 by definition; 3 fields
were not tested. These empirical weights are a post-hoc measurement of
effect sizes: the benchmark's retained scoring retains the prespecified
weights, and we do not evaluate a scorer reweighted by these effects
against reproduction outcomes in this work, which would require the
campaign of Section 3.8 and would otherwise risk circularity.

\subsection{Executability strata as a routing
signal}\label{executability-strata-as-a-routing-signal}

Executability class indicates whether re-execution is possible from the
reported information and contributes to the feasibility term
\(f_{\text{feas}}(E)\). It is not interpreted as a probability of
reproduction. Human verification supported reporting-status extraction,
but agreement on the ordinal E-class was moderate (linear-weighted
\(\kappa=0.47\)). Among 37 quality-control-passed re-executions, 78\%
were within tolerance. All 6 E3 claims were within tolerance, compared
with 23 of 31 E2 claims, but this comparison was underpowered and
confounded by box reporting (Mann-Whitney \(P=0.138\)). The continuous
quality score was also unrelated to reproduction error. We therefore use
E-class as a routing signal that indicates when a claim is directly
re-executable, not as a predictor of whether it reproduces. The
framework is restricted to computational re-execution of box-based
docking and does not address wet-lab replication or boxless methods such
as DiffDock (Corso et al. 2023).

\section{Methods}\label{methods}

We evaluated reporting criteria using an audit of reporting
completeness, controlled one-factor perturbations with two measured
noise floors, and re-execution of reported literature claims. We then
derived empirical field weights from the perturbation results.
Statistical resampling used seed 2026; ligand generation and the
principal docking protocol used seed 200. Analyses consumed retained
outputs for offline verification (Sandve et al. 2013; Stodden et al.
2016; Peng 2011).

\subsection{Audit corpus and span-verified
extraction}\label{audit-corpus-and-span-verified-extraction}

The audit corpus comprised 50 open-access molecular docking papers
spanning multiple target topics. A large language model extracted 12
reporting fields: docking engine, PDB receptor, protein chain, receptor
preparation, ligand preparation, software version, grid centre, grid
size, exhaustiveness, random seed, redocking validation and numeric top
affinity. We accepted an extracted value only when it matched a verbatim
span in the source text (Dagdelen et al. 2024; {Gartlehner et al.} 2024;
{Konet et al.} 2024; {Yisha et al.} 2026). This was a reporting audit
rather than a systematic review; PRISMA 2020 was used as a reference for
transparent selection reporting ({Page et al.} 2021) and summarised
completeness with verified counts and Wilson 95\% confidence intervals
(Table 1). One developer-reviewer manually adjudicated a
topic-stratified 30\% subset comprising 15 papers, 180 field judgements
and 15 E-class judgements. Gemini and GPT outputs provided model
cross-checks, but the developer-reviewer assigned every final label. We
analysed exact-value and reporting-status agreement separately,
summarised reporting-status reliability with Cohen's kappa and the
prevalence-robust Gwet AC1, and assessed ordinal E-class agreement with
linear-weighted kappa. Reporting-status agreement is the audit-relevant
quantity because completeness is a reported-versus-not decision;
exact-value disagreements need not change whether a field is present.
Because one human established the reference labels, these coefficients
describe expert-versus-model agreement rather than independent human
inter-rater reliability; no second human coder was used.

\subsection{Controlled re-execution
ablation}\label{controlled-re-execution-ablation}

The perturbation set was drawn from RCSB Protein Data Bank X-ray
complexes (Burley et al. 2018) with crystallographic resolution below
1.8 \AA{}, an experimentally measured \(K_d\) or \(K_i\), and a single
crystallographic ligand that defined the docking box. RDKit ETKDGv3
generated ligand three-dimensional coordinates with random seed 200
(Landrum et al. 2026), Meeko prepared receptors and ligands
(Santos-Martins et al. 2025), and AutoDock Vina 1.2.5 performed docking
at exhaustiveness 16 (Trott and Olson 2010; Eberhardt et al. 2021). We
changed one parameter at a time while holding the remaining inputs fixed
and recorded the change in top-pose score (Table 2).

The perturbations were box size (tightened to a 25 \AA{} cube),
exhaustiveness (reduced from 16 to 4), box centre (a blind whole-protein
box versus the site-centred box), random seed, protonation state, ligand
identity and receptor structure. We tested protonation by re-protonating
and re-docking each ligand. The primary protonation summary used the 37
comparison rows tagged ionizable in the retained decomposition table;
seven non-ionizable rows were excluded from that summary, following the
retained analysis code. We tested ligand identity by docking plausible
tautomeric and stereoisomeric variants of ambiguously named ligands,
reporting both the absolute variant-minus-canonical shift and the
per-ligand ambiguity spread. We tested receptor structure by
cross-docking each ligand into the other high-resolution structures of
its target (3 targets, 6 structures each, 85 non-cognate comparisons).
The resulting shift measured score dispersion attributable to
receptor-conformer choice rather than a directional induced-fit penalty.
We additionally confirmed the box-size and exhaustiveness perturbations
in a diverse 44-complex set under uniform local preparation. Because
cognate structures did not systematically produce the most favourable
score, we interpreted the cross-docking effect as dispersion rather than
the near-uniform redocking success reported for cognate complexes
elsewhere (Flachsenberg et al. 2024; {Zajáček et al.} 2024; Francoeur et
al. 2020).

Ligand identifiability, a categorical prerequisite upstream of every
graded parameter, was measured as the fraction of unique reported ligand
names in the reproduction corpus (n = 116) that could not be resolved to
a chemical structure by a name-to-PubChem-CID lookup (Kim et al. 2025).

\subsection{Noise floors and the two-floor reproduction
tolerance}\label{noise-floors-and-the-two-floor-reproduction-tolerance}

Each recorded shift was interpreted against two empirically measured
noise floors and a reproduction tolerance. The within-machine
self-consistency floor was estimated from repeated docking of 15 ligands
across 3 random seeds, giving a median run-to-run range of 0.04 kcal/mol
(\texttt{reproduction\_LOCKED.json}); this summarizes stochastic score
variation in the tested runs. The cross-environment floor was estimated
by re-executing the same manifest on independent cloud virtual machines
and comparing against the local run. Two independent re-executions gave
median absolute differences of 0.43 kcal/mol (a more divergent RDKit
build, n = 15) and 0.18 kcal/mol (a nearer software stack, n = 35;
Section 4.7), so the measured cross-environment variation is
environment-pair-dependent. We use 0.43 kcal/mol as a conservative
comparison reference for this study, not a universal lower bound on
distinguishable effects or calibrated probabilities. The reproduction
tolerance was set at 2.0 kcal/mol. These three quantities are strictly
ordered, and the reproduction outcome \(Y_j\) is a thresholded readout
of a rerun affinity landing within tolerance of the reported value:

\[
\tau_{\text{self}} = 0.04 \ \leq\ \tau_{\text{cross-env}} = 0.43 \ \ll\ \tau_{\text{repro}} = 2.0 \ \text{kcal/mol},
\]

\[
Y_j = \mathbf{1}\!\left[\,\bigl|\,a_{\text{rep}} - a_{\text{rerun}}\,\bigr| \le \tau_{\text{repro}}\,\right],\qquad \tau_{\text{repro}} > \tau_{\text{cross-env}} > \tau_{\text{self}}.
\]

Because the 2.0 kcal mol\(^{-1}\) tolerance exceeded both measured
floors, \(Y\) primarily captured differences larger than run-to-run or
cross-environment variation. Cross-environment variation should be
evaluated when interpreting claims near the reproduction cutoff.

\subsection{Validation with Vinardo}\label{validation-with-vinardo}

To test whether the ordering of input effects depended on the Vina
scoring terms, we repeated the analysis using the same prepared PDBQT
receptors and ligands, box centres and sizes, random seed 200 and
exhaustiveness 16, changing only the scoring function to Vinardo
(Quiroga and Villarreal 2016). A Vina positive control reproduced the
recorded cross-dock score exactly. We quantified agreement in
per-structure receptor effects using Spearman correlation.

\subsection{Generalization: target
diversity}\label{generalization-target-diversity}

To assess generalisation beyond the original rigid enzymes, we applied
the same cross-docking procedure to 9 additional targets from 6
additional protein families: CDK2, MAPK14, ABL1, MAPK1, Factor Xa,
BACE1, HSP90, ESR1 and PDE5A. The complete benchmark therefore included
12 targets from 7 protein families. Of 305 attempted docking runs, 238
succeeded and yielded 191 valid non-cognate cross-self comparisons
(Table 3). Failed runs remained in the run ledger and were not imputed.
We selected RCSB X-ray structures with resolution below 2.2 \AA{} and a
distinct cognate ligand for each structure. The cognate-ligand centroid
and extent defined the docking box. This block used random seed 200,
exhaustiveness 8 and a maximum box length of 24 \AA{} on a Google Cloud
CPU virtual machine. The preceding perturbation analysis showed that
changes in box size and exhaustiveness were small relative to the
cross-environment floor, supporting these reduced search settings for
the receptor-structure analysis.

\subsection{Reproduction harness (N=37)}\label{reproduction-harness-n37}

We re-executed reported docking claims using AutoDock Vina 1.2.7,
reported box sizes capped at 40 \AA{}, RDKit ETKDGv3 with random seed
200 and Meeko preparation (Landrum et al. 2026; Santos-Martins et al.
2025). A deterministic pipeline attempted to resolve each reported
ligand name or identifier to a PubChem compound identifier and then to
an InChIKey and canonical SMILES (Kim et al. 2025). A ligand string
passed the automated gate if the paper supplied an explicit identifier
or if exact and normalized name queries returned one unique PubChem CID;
queries returning no CID or multiple CIDs remained unresolved or
ambiguous. PubChem responses were cached, and transient HTTP failures
stopped the analysis rather than being counted as chemical
non-resolution. This structural anchoring reduced name drift; unresolved
strings were recorded rather than guessed. The pipeline fetched each
target structure, removed waters, ions and the co-crystallized ligand,
and converted the protonated receptor to PDBQT. It used reported
docking-box coordinates when available; otherwise, it centred the box on
the co-crystal ligand and flagged the claim as assumption-dependent, so
an assumed box was never counted as fully reported. Every claim
generated a provenance record containing the resolved SMILES, compound
identifier, box, seed, exhaustiveness, engine version and exact command,
allowing the execution request to be reconstructed exactly (Groth and
Moreau 2013; RO-Crate Community, n.d.). Three of the 40 obtained local
docking results met the retained exclusion criterion (rerun score above
-2 kcal mol\(^{-1}\), interpreted as no bound pose), leaving 37
quality-control-passed claims from 7 papers (Table 5). These rates are
conditional on obtaining a docking result and passing this QC rule;
unresolved or failed manifest entries do not enter the within-tolerance
denominator. Most claims concerned the SARS-CoV-2 main protease
structures 6LU7 and 6Y2E, drawn from published main-protease docking and
virtual-screening studies ({Ambrosio et al.} 2023; {Mandour et al.}
2022; {Peralta-Moreno et al.} 2023; {Sisakht et al.} 2021). We defined
reproduction within tolerance as an absolute reported-rerun difference
of 2.0 kcal mol\(^{-1}\) or less and stratified claims by their pre-run
executability class. Only 6 claims were classified as E3, and E-class
was confounded with box reporting; comparisons between E3 and E2 were
therefore treated as descriptive.

\subsection{Statistics}\label{statistics}

We assessed whether search-space effects were practically negligible
using a one-sided equivalence test with the 0.43 kcal mol\(^{-1}\)
cross-environment floor as the margin, together with a
margin-sensitivity analysis. We calculated Wilson score 95\% confidence
intervals for reporting completeness and reproduction proportions,

\begin{equation}
\mathrm{CI}=\frac{\hat{p}+\frac{z^2}{2n}\ \pm\ z\sqrt{\dfrac{\hat{p}(1-\hat{p})}{n}+\dfrac{z^2}{4n^2}}}{1+\frac{z^2}{n}},\qquad z=1.96,
\end{equation}

which is better behaved than the normal approximation for the small
counts and near-boundary proportions in the audit, and a paper-clustered
bootstrap interval for the overall reproduction proportion. We used
Spearman correlation for rank agreement across scoring functions and
parameters, Mann-Whitney U tests for between-group comparisons, and
Cronbach's alpha for internal consistency of the reporting subscale
(Cronbach 1951; Tavakol and Dennick 2011). We controlled multiple
comparisons using the Benjamini-Hochberg false discovery rate.
Statistical resampling used seed 2026; docking used the
protocol-specific seed reported above. Multiple comparisons sharing a
ligand or receptor are not independent. Pooled comparison tests and
their Benjamini-Hochberg adjustments are exploratory; the ligand-level
receptor analysis provides a complementary summary without treating each
cross-dock pair as an independent ligand.

\subsection{Powered design for a calibration
campaign}\label{powered-design-for-a-calibration-campaign}

The calibration map of Section 2.7 and the executability-stratum
contrast of Section 4.6 are not estimable at the present scale, and we
sized a forward re-execution campaign for each inferential goal. For the
E3-versus-E2 reproduction contrast, a two-proportion test with
anticipated proportions \(p_1=0.90\) and \(p_2=0.60\), power 0.8 and
two-sided \(\alpha=0.05\) requires

\begin{equation}
\frac{n}{\text{group}}=\frac{(z_{\alpha/2}+z_\beta)^2\,\big[p_1(1-p_1)+p_2(1-p_2)\big]}{(p_1-p_2)^2}\approx 29,
\end{equation}

which the paper-level clustering inflates by the design effect

\begin{equation}
\mathrm{DEFF}=1+(m-1)\,\mathrm{ICC},
\end{equation}

giving approximately 40 claims per stratum at \(m\approx5\) claims per
paper and an intraclass correlation of 0.1; the present \(n_{E3}=6\) is
accordingly far below the powered target, consistent with the observed
\(P=0.14\). Fitting the calibration map of Section 2.7 requires roughly
10 events per bin across at least 15 paper clusters, or about 100 to 150
quality-control-passed re-executions, and we propose about 250 audited
records as a planning target for an item-response model of the audit
scale (Section 2.4), with adequacy dependent on item prevalence and
model fit, which is an inexpensive reading task that needs no docking
compute. The binding target for a re-execution campaign is therefore
about 150 quality-control-passed re-executions distributed over at least
15 papers and at least three protein families, with at least 30 to 40
claims per executability stratum. None of the relevance threshold, the
core-subscale weights or the reproduction tolerance may be tuned to
reproduction outcomes; a future campaign would need a public
analysis-plan registration and executable checks enforcing this
constraint; the campaign described here is a planning proposal (Sandve
et al. 2013; {Rule et al.} 2019).

\section{Results}\label{results}

\subsection{Reporting completeness differed among method
fields}\label{reporting-completeness-differed-among-method-fields}

Reporting completeness varied substantially among the 12 audited fields
(Table 1). The docking engine was reported in all 50 papers (100\%;
Wilson 95\% confidence interval (CI), 93--100\%), and the receptor
structure was reported in 47 (94\%; 95\% CI, 84--98\%). Receptor
preparation, numeric top affinity and ligand preparation were reported
in 86\%, 80\% and 72\% of papers, respectively. By contrast, grid size
was reported in 42\%, grid centre in 34\%, exhaustiveness in 20\%,
protein chain in 18\% and random seed in 2\% (1/50; 95\% CI, 0--10\%).
The fields that were reliably reported identify the calculation at a
high level, namely the engine, the target and the numeric result,
whereas the fields that specify the search box and the stochastic
settings were the ones most often omitted. The random seed, which fixes
the stochastic component of the search, was reported in a single paper,
so a reader attempting to reproduce the remaining 49 calculations
exactly would have to assume it.

In the 15-paper manually adjudicated subset, model-extracted and
human-adjudicated field values agreed in 143 of 180 judgements (79.4\%).
Reporting status, which is the quantity the completeness audit actually
measures, agreed in 169 of 180 judgements (93.9\%; Cohen's
\(\kappa=0.87\), Gwet's AC1\(=0.88\)), exceeding the prespecified
threshold of 0.6. Several fields were nearly universal or nearly absent,
a prevalence pattern that can deflate chance-corrected kappa despite
high raw agreement. We therefore report Gwet's AC1 as a
prevalence-robust companion coefficient; the two coefficients agreed.
Several disagreements concerned the extracted value rather than field
presence, explaining the lower value agreement. E-class agreement was
lower (6/15 exact; linear-weighted \(\kappa=0.47\)), so we retained
E-class as a descriptive executability stratum. Every accepted model
extraction matched a source-text span (Dagdelen et al. 2024; {Gartlehner
et al.} 2024).

\textbf{Table 1. Reporting completeness of 12 docking method fields in
the 50-paper open-access corpus. Values are verified counts, percentages
and Wilson 95\% confidence intervals. Fields are ordered from most to
least frequently reported.}

{\def\LTcaptype{none} 
\begin{longtable}[]{@{}
  >{\raggedright\arraybackslash}p{(\linewidth - 8\tabcolsep) * \real{0.2000}}
  >{\raggedright\arraybackslash}p{(\linewidth - 8\tabcolsep) * \real{0.2000}}
  >{\raggedright\arraybackslash}p{(\linewidth - 8\tabcolsep) * \real{0.2000}}
  >{\raggedright\arraybackslash}p{(\linewidth - 8\tabcolsep) * \real{0.2000}}
  >{\raggedright\arraybackslash}p{(\linewidth - 8\tabcolsep) * \real{0.2000}}@{}}
\toprule\noalign{}
\begin{minipage}[b]{\linewidth}\raggedright
Field
\end{minipage} & \begin{minipage}[b]{\linewidth}\raggedright
Verified n / N
\end{minipage} & \begin{minipage}[b]{\linewidth}\raggedright
Reported \%
\end{minipage} & \begin{minipage}[b]{\linewidth}\raggedright
Wilson 95\% CI
\end{minipage} & \begin{minipage}[b]{\linewidth}\raggedright
Tier (reporting)
\end{minipage} \\
\midrule\noalign{}
\endhead
\bottomrule\noalign{}
\endlastfoot
Docking software (engine) & 50/50 & 100\% & 93--100\% & Near-universal
(\textgreater90\%) \\
PDB receptor (target structure) & 47/50 & 94\% & 84--98\% &
Near-universal (\textgreater90\%) \\
Receptor preparation & 43/50 & 86\% & 74--93\% & Well reported
(60--90\%) \\
Numeric result (top affinity) & 40/50 & 80\% & 67--89\% & Well reported
(60--90\%) \\
Ligand preparation & 36/50 & 72\% & 58--83\% & Well reported
(60--90\%) \\
Software version & 32/50 & 64\% & 50--76\% & Well reported (60--90\%) \\
Redocking validation & 24/50 & 48\% & 35--61\% & Sparse (20--50\%) \\
Grid size & 21/50 & 42\% & 29--56\% & Sparse (20--50\%) \\
Grid centre & 17/50 & 34\% & 22--48\% & Sparse (20--50\%) \\
Exhaustiveness & 10/50 & 20\% & 11--33\% & Sparse (20--50\%) \\
Protein chain & 9/50 & 18\% & 10--31\% & Rare (\textless20\%) \\
Random seed & 1/50 & 2\% & 0--10\% & Rare (\textless20\%) \\
\end{longtable}
}

\subsection{Search-space perturbations produced small score
changes}\label{search-space-perturbations-produced-small-score-changes}

We next tested whether the sparsely reported search fields materially
affected the reproduced score (Table 2, Fig. 2). Each crystallographic
ligand was redocked from a fully specified baseline in AutoDock Vina
(Trott and Olson 2010), with one parameter changed at a time. We
interpreted the resulting score changes relative to two measured noise
floors and a reproduction tolerance:

\begin{figure}
\centering
\includegraphics[width=1\linewidth,height=\textheight,keepaspectratio,alt={Input-specific effects on re-executed docking scores. (a) Individual absolute score changes, interquartile ranges and medians for search specification, ligand chemistry and receptor conformation; dotted and dashed lines mark the 0.04 kcal mol\^{}\{-1\} self-consistency and 0.43 kcal mol\^{}\{-1\} cross-environment floors. (b) Fraction of perturbations at or above each effect threshold. (c-e) Absolute cross-self score changes for the original carbonic anhydrase II (CAII), thrombin (THR) and trypsin (TRYP) receptor ensembles. Rows are ligands, columns are receptor conformers and open circles identify cognate ligand-structure pairs.}]{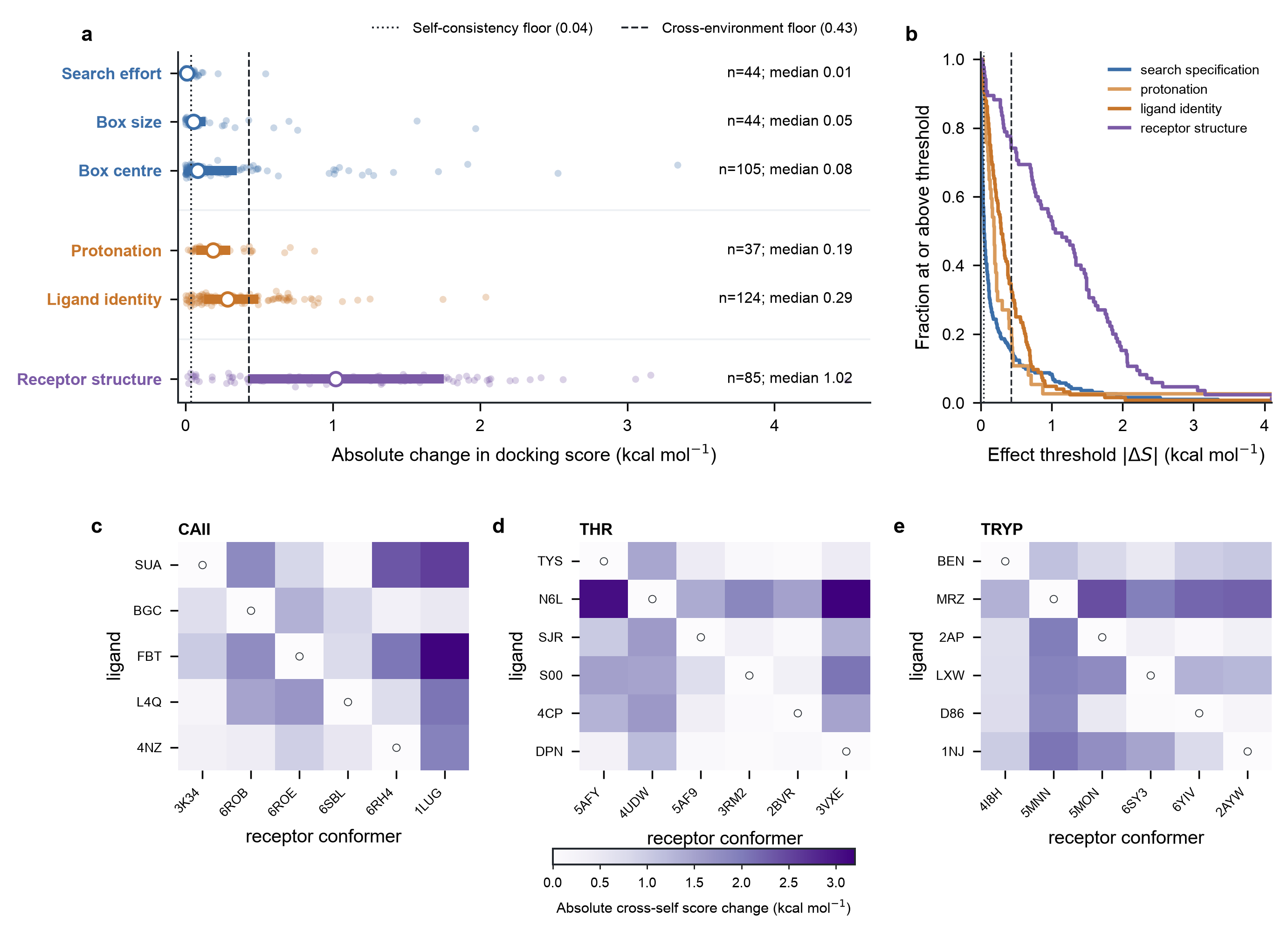}
\caption{Input-specific effects on re-executed docking scores. (a)
Individual absolute score changes, interquartile ranges and medians for
search specification, ligand chemistry and receptor conformation; dotted
and dashed lines mark the 0.04 kcal mol\(^{-1}\) self-consistency and
0.43 kcal mol\(^{-1}\) cross-environment floors. (b) Fraction of
perturbations at or above each effect threshold. (c-e) Absolute
cross-self score changes for the original carbonic anhydrase II (CAII),
thrombin (THR) and trypsin (TRYP) receptor ensembles. Rows are ligands,
columns are receptor conformers and open circles identify cognate
ligand-structure pairs.}
\end{figure}

\[\tau_{\text{self}} = 0.04 \ \leq\ \tau_{\text{cross-env}} = 0.43 \ \ll\ \tau_{\text{repro}} = 2.0 \ \text{kcal/mol}, \qquad Y_j = \mathbf{1}\!\left[\ \big|a_{\text{rep}} - a_{\text{rerun}}\big| \leq \tau_{\text{repro}}\right],\]

The within-machine floor was 0.04 kcal mol\(^{-1}\), estimated from 15
ligands across 3 random seeds, and the cross-environment floor was 0.43
kcal mol\(^{-1}\). Box size changed the score by a median of 0.054 kcal
mol\(^{-1}\) (n = 44), reducing exhaustiveness from 16 to 4 changed it
by 0.010 (n = 44), and replacing the site-centred box with a blind box
changed it by 0.084 (n = 105). Random-seed variation equalled the 0.04
kcal mol\(^{-1}\) self-consistency floor (n = 15). All 4 effects were
equivalent to negligible using the 0.43 kcal mol\(^{-1}\) margin; the
Supplement reports the margin-sensitivity analysis. Thus, under the
tested Vina protocol, search-space perturbations contributed little
relative to receptor and ligand inputs.

\textbf{Table 2. Effects of one-factor perturbations on the reproduced
docking score, ordered by the median absolute score change
(\(|\Delta S|\), kcal mol\(^{-1}\)). Search-space perturbations produced
median changes of 0.08 kcal mol\(^{-1}\) or less, whereas receptor
structure produced the largest change. The Vinardo column reports
repeated analyses for the upper and lower ends of this ordering;
intermediate fields were not retested with Vinardo. Pooled per-structure
effects were correlated between scoring functions (the nominal test
treats 85 pairs as independent despite shared ligands and receptors)
(Spearman \(\rho = 0.64\), \(P = 4.5 \times 10^{-11}\), n = 85).}

{\def\LTcaptype{none} 
\begin{longtable}[]{@{}
  >{\raggedright\arraybackslash}p{(\linewidth - 8\tabcolsep) * \real{0.2000}}
  >{\raggedright\arraybackslash}p{(\linewidth - 8\tabcolsep) * \real{0.2000}}
  >{\raggedright\arraybackslash}p{(\linewidth - 8\tabcolsep) * \real{0.2000}}
  >{\raggedright\arraybackslash}p{(\linewidth - 8\tabcolsep) * \real{0.2000}}
  >{\raggedright\arraybackslash}p{(\linewidth - 8\tabcolsep) * \real{0.2000}}@{}}
\toprule\noalign{}
\begin{minipage}[b]{\linewidth}\raggedright
Determinant (parameter)
\end{minipage} & \begin{minipage}[b]{\linewidth}\raggedright
Vina median \textbar dS\textbar{}
\end{minipage} & \begin{minipage}[b]{\linewidth}\raggedright
n
\end{minipage} & \begin{minipage}[b]{\linewidth}\raggedright
Within-tolerance (\textless=2.0)
\end{minipage} & \begin{minipage}[b]{\linewidth}\raggedright
Vinardo median \textbar dS\textbar{}
\end{minipage} \\
\midrule\noalign{}
\endhead
\bottomrule\noalign{}
\endlastfoot
Receptor structure (cross-dock, pooled) & 1.02 & 85 & 73/85 (86\%) &
0.96 \\
Receptor structure (ligand-level) & 1.30 & 17 & not applicable & not
re-run \\
Ligand identity (pairwise variant) & 0.287 & 124 & 123/124 (99\%) & not
re-run \\
Ligand identity (worst-case spread) & 0.580 & 28 & not applicable & not
re-run \\
Protonation & 0.188 & 37 & 37/37 (100\%) & not re-run \\
Box center (blind vs site-centred) & 0.084 & 105 & 103/105 (98\%) & not
re-run \\
Box size (tight 25\textsuperscript{3}) & 0.054 & 44 & 44/44 (100\%) &
0.082 \\
Exhaustiveness (16\ensuremath{\rightarrow}4) & 0.010 & 44 & 44/44
(100\%) & 0.008 \\
Random seed (self-consistency floor) & 0.04 & 15 & 15/15 (100\%) & not
re-run \\
\end{longtable}
}

\subsection{Receptor structure produced the largest effect across
scoring functions and
targets}\label{receptor-structure-produced-the-largest-effect-across-scoring-functions-and-targets}

Physical input changes produced larger effects (Table 2, Fig. 2).
Changing ligand protonation altered the score by a median of 0.188 kcal
mol\(^{-1}\) (n = 37) (Brink and Exner 2009). Alternative tautomeric or
stereoisomeric ligand identities produced a median pairwise change of
0.287 kcal mol\(^{-1}\) (n = 124) and a worst-case spread of 0.580 kcal
mol\(^{-1}\) (n = 28). Receptor structure produced the largest effect:
non-cognate cross-docking changed the score by a pooled median of 1.02
kcal mol\(^{-1}\) and a ligand-level median of 1.30 kcal mol\(^{-1}\) (n
= 85; ligand-level Wilcoxon \(P<0.001\)). The cognate structure gave the
most favourable score for only 3 of 17 ligands, indicating dispersion
across receptor conformations rather than a uniformly directional
penalty. Nevertheless, 73 of 85 comparisons (86\%) remained within the
2.0 kcal mol\(^{-1}\) tolerance.

The ordering was retained with Vinardo (Quiroga and Villarreal 2016)
(Table 2, Fig. 3a). The median receptor-structure effect was 0.955 kcal
mol\(^{-1}\), whereas the effects of box size and exhaustiveness were
0.082 and 0.008 kcal mol\(^{-1}\), respectively. Per-structure effects
correlated across the two scoring functions (Spearman \(\rho=0.64\),
\(P=4.5\times10^{-11}\), n = 85). Across 9 additional targets, 191 valid
non-cognate comparisons yielded a pooled receptor-structure effect of
0.84 kcal mol\(^{-1}\), which did not differ from the 1.02 kcal
mol\(^{-1}\) value in the original enzymes (Mann-Whitney \(P=0.48\);
Table 3, Fig. 3b). ABL1 and MAPK14 had the largest target-level median
effects (2.23 and 1.99 kcal mol\(^{-1}\)), whereas PDE5A and Factor Xa
had the smallest (0.38 and 0.61 kcal mol\(^{-1}\)), consistent with
target-dependent conformational sensitivity in cross-docking (Francoeur
et al. 2020).

\begin{figure}
\centering
\includegraphics[width=1\linewidth,height=\textheight,keepaspectratio,alt={Generalization of receptor-structure effects. (a) Paired Vina and Vinardo effects for 85 non-cognate comparisons; the dashed line is identity. (b) Effect distributions for the two scoring functions. (c) Target-level median effects among 9 additional targets, coloured by protein family. Per-structure effects correlated across scoring functions (Spearman \textbackslash rho = 0.64, P = 4.5 \textbackslash times 10\^{}\{-11\}).}]{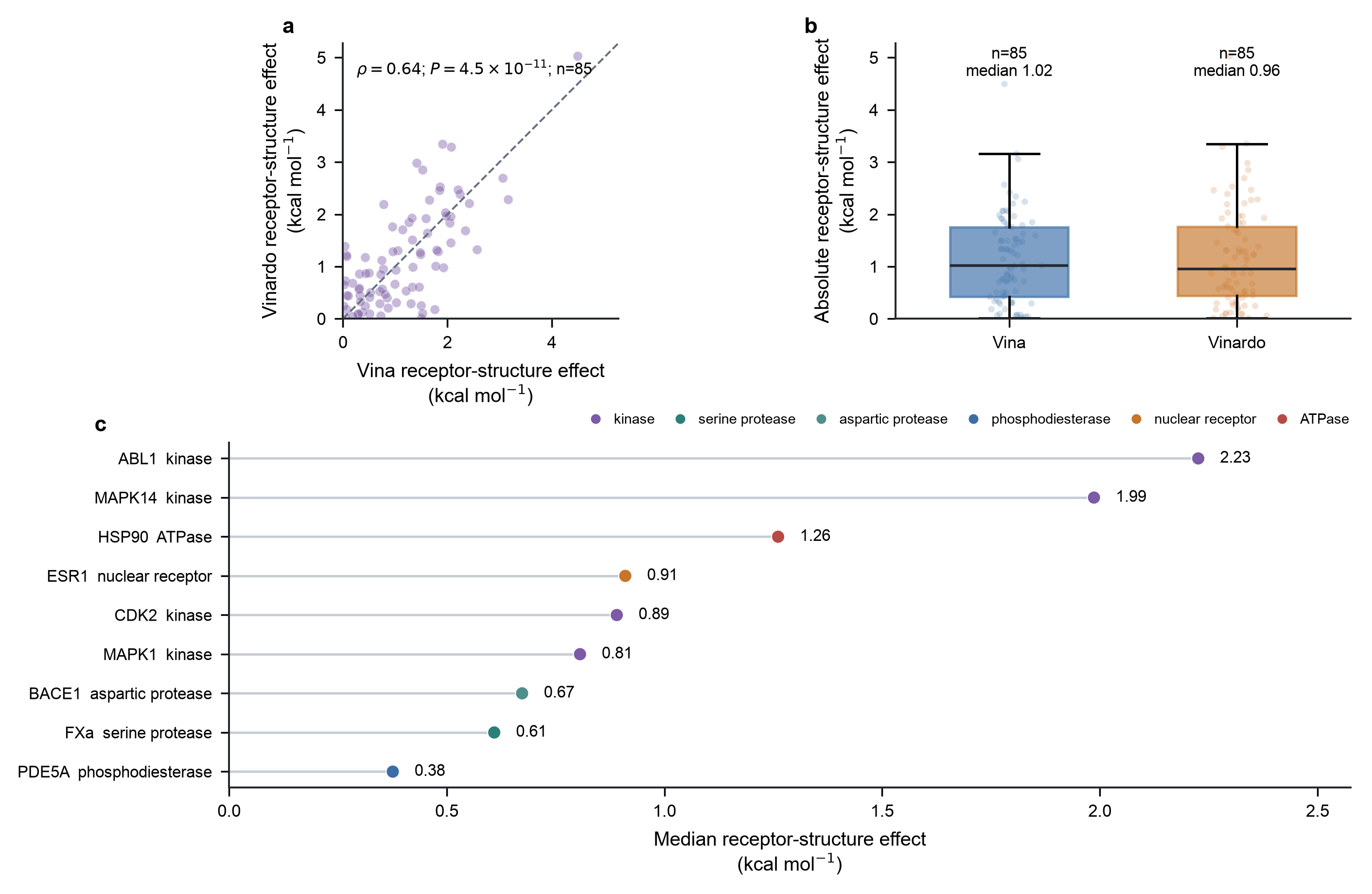}
\caption{Generalization of receptor-structure effects. (a) Paired Vina
and Vinardo effects for 85 non-cognate comparisons; the dashed line is
identity. (b) Effect distributions for the two scoring functions. (c)
Target-level median effects among 9 additional targets, coloured by
protein family. Per-structure effects correlated across scoring
functions (Spearman \(\rho = 0.64\), \(P = 4.5 \times 10^{-11}\)).}
\end{figure}

\textbf{Table 3. Receptor-structure effects across 12 targets in 7
protein families. Values are per-target median absolute cross-self score
changes (kcal mol\(^{-1}\)), numbers of successful non-cognate
comparisons and proportions within the 2.0 kcal mol\(^{-1}\) tolerance.
The original 3-target set had a pooled median of 1.02 kcal mol\(^{-1}\)
(n = 85), and the 9 additional targets had a pooled median of 0.84 kcal
mol\(^{-1}\) (n = 191; Mann-Whitney \(P = 0.48\)). Of 305 attempted runs
in the additional-target block, 238 succeeded and yielded 191 valid
non-cognate comparisons. Targets are ordered by effect within each
block.}

{\def\LTcaptype{none} 
\begin{longtable}[]{@{}lllll@{}}
\toprule\noalign{}
Target & Family & Median \(|\Delta S|\) & n & Within 2.0 \\
\midrule\noalign{}
\endhead
\bottomrule\noalign{}
\endlastfoot
\textbf{Rigid classical enzymes (original)} & & & & \\
THR (thrombin) & Serine protease & 1.12 & 30 & 90\% \\
TRYP (trypsin) & Serine protease & 1.10 & 30 & 87\% \\
CAII (carbonic anhydrase) & Lyase (metalloenzyme) & 0.95 & 25 & 80\% \\
Rigid POOLED & 3 targets & 1.02 & 85 & 86\% \\
\textbf{Diverse flexible targets} & & & & \\
ABL1 & Kinase & 2.23 & 9 & 44\% \\
MAPK14 (p38) & Kinase & 1.99 & 20 & 50\% \\
HSP90 & ATPase (chaperone) & 1.26 & 25 & 68\% \\
ESR1 & Nuclear receptor & 0.91 & 30 & 100\% \\
CDK2 & Kinase & 0.89 & 30 & 80\% \\
MAPK1 (ERK2) & Kinase & 0.81 & 15 & 100\% \\
BACE1 & Aspartic protease & 0.67 & 30 & 97\% \\
FXa (Factor Xa) & Serine protease & 0.61 & 19 & 79\% \\
PDE5A & Phosphodiesterase & 0.38 & 13 & 100\% \\
Diverse POOLED & 9 targets & 0.84 & 191 & 82\% \\
\end{longtable}
}

\subsection{Unresolved or ambiguous ligand strings blocked automated
re-execution}\label{unresolved-or-ambiguous-ligand-strings-blocked-automated-re-execution}

Of 116 unique reported ligand strings in the reproduction corpus, 26
(22\%; Wilson 95\% CI, 16--31\%) remained unresolved or ambiguous after
exact and deterministically normalized name-to-CID queries (Fig. 4b).
These records commonly used series labels such as ``compound 5'', ``N3''
or ``imidazolyl-methanone C10'' without a structural identifier. They
did not enter automated re-execution and were therefore treated as
failing a prerequisite rather than receiving a graded perturbation
score. This proportion characterizes the prespecified resolver applied
to the reported strings; it does not imply that manual chemical curation
could never identify a structure.

\begin{figure}
\centering
\includegraphics[width=1\linewidth,height=\textheight,keepaspectratio,alt={Reporting-effect inversion, ligand resolvability and empirical reordering. (a) Reporting frequency versus measured effect for the 5 fields shared by the 50-paper reporting audit and direct perturbation design; dotted and dashed lines mark the computational floors. (b) Resolution of 116 unique reported ligand names to machine-readable structures: 41 had explicit identifiers, 49 resolved to one PubChem CID after deterministic name normalization and 26 remained unresolved or ambiguous. (c) Rank transitions for the 7 directly perturbed fields from prespecified reporting tier to measured median absolute score change. Gate denotes the ligand-resolvability prerequisite.}]{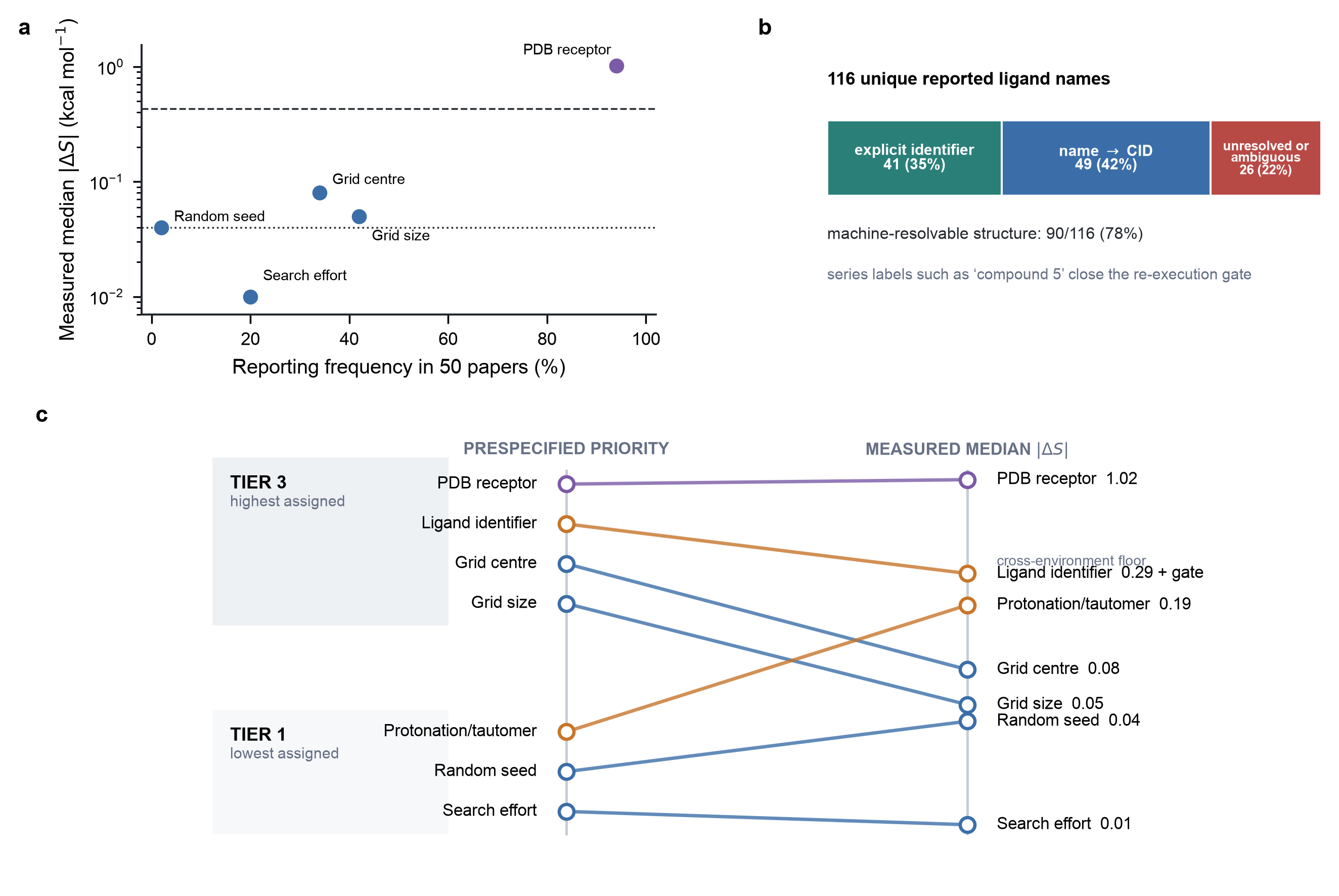}
\caption{Reporting-effect inversion, ligand resolvability and empirical
reordering. (a) Reporting frequency versus measured effect for the 5
fields shared by the 50-paper reporting audit and direct perturbation
design; dotted and dashed lines mark the computational floors. (b)
Resolution of 116 unique reported ligand names to machine-readable
structures: 41 had explicit identifiers, 49 resolved to one PubChem CID
after deterministic name normalization and 26 remained unresolved or
ambiguous. (c) Rank transitions for the 7 directly perturbed fields from
prespecified reporting tier to measured median absolute score change.
\protect\texttt{Gate} denotes the ligand-resolvability prerequisite.}
\end{figure}

\subsection{Empirical effects reordered prespecified field
priorities}\label{empirical-effects-reordered-prespecified-field-priorities}

Setting each method-field weight to a monotone function of its measured
perturbation effect, capped at the reproduction tolerance,

\[w_{\text{field}} = \min\!\big(\operatorname{median}\big|\Delta S_{\text{field}}\big|,\ \tau\big),\]

\[\begin{aligned} w_{\text{pdb}}=1.02 &> w_{\text{ligand\_id}}=0.29\,(+\text{gate}) > w_{\text{protonation}}=0.19 > w_{\text{grid\_centre}}=0.08 \\ &> w_{\text{grid\_size}}=0.05 > w_{\text{seed}}=0.04 > w_{\text{exh}}=0.01. \end{aligned}\]

reordered the prespecified priorities (Table 4, Fig. 4a,c). Grid centre
and grid size were assigned to tier 3 in the prespecified framework but
had empirical effects of 0.08 and 0.05 kcal mol\(^{-1}\), respectively.
Protonation was assigned to tier 1 but had a larger effect of 0.19 kcal
mol\(^{-1}\). Receptor structure had the largest empirical weight
(1.02), and ligand identity combined a 0.29 kcal mol\(^{-1}\) effect
with the 22\% unresolved-or-ambiguous prerequisite. These values
constitute a post-hoc analysis result.

The evidence supporting the 16 fields also differed (Fig. 5). Seven
fields were tested directly, 2 were supported by partial perturbations,
3 were characterised using reproduction outcomes and 1 was defined by
the benchmark scope. Receptor preparation, water and ion handling, and
code-artifact availability were not tested. This classification
distinguishes measured perturbation effects from outcome associations
and untested assumptions.

\begin{figure}
\centering
\includegraphics[width=0.92\linewidth,height=\textheight,keepaspectratio,alt={Evidence grounding and internal validation of the NSE score. (a) Representative carbonic anhydrase II structure (PDB 4YX4) with the FB2 ligand and catalytic zinc, annotated with the largest measured input effects (Gaspari et al. 2016). (b) Exact-value and reported-versus-missing agreement between model extraction and final developer adjudication for 12 fields in 15 papers; the dotted line marks 0.80 agreement. (c) Span-verified 12-field reporting matrix for 50 papers ordered by completeness. (d) Evidence source for all 16 NSE fields: direct perturbation, partial perturbation, re-execution outcome, benchmark definition or not tested. (e) Off-topic decoy rejection: best-evidence scores for 33 on-topic real papers and 6 injected off-topic decoys under the hard-gated aggregator B and an additive aggregator; the hard gate assigns every decoy a zero score (area under the receiver operating characteristic curve 1.00), whereas the additive aggregator lets decoys leak into the real-paper range (0.89). (f) Reliability of the reporting-readiness subscale: item-rest discriminations of the five-item core subscale, which has Cronbach \textbackslash alpha = 0.73 (full 12-item \textbackslash alpha = 0.62; seven fields were retained as descriptive checklist items outside this selected subscale).}]{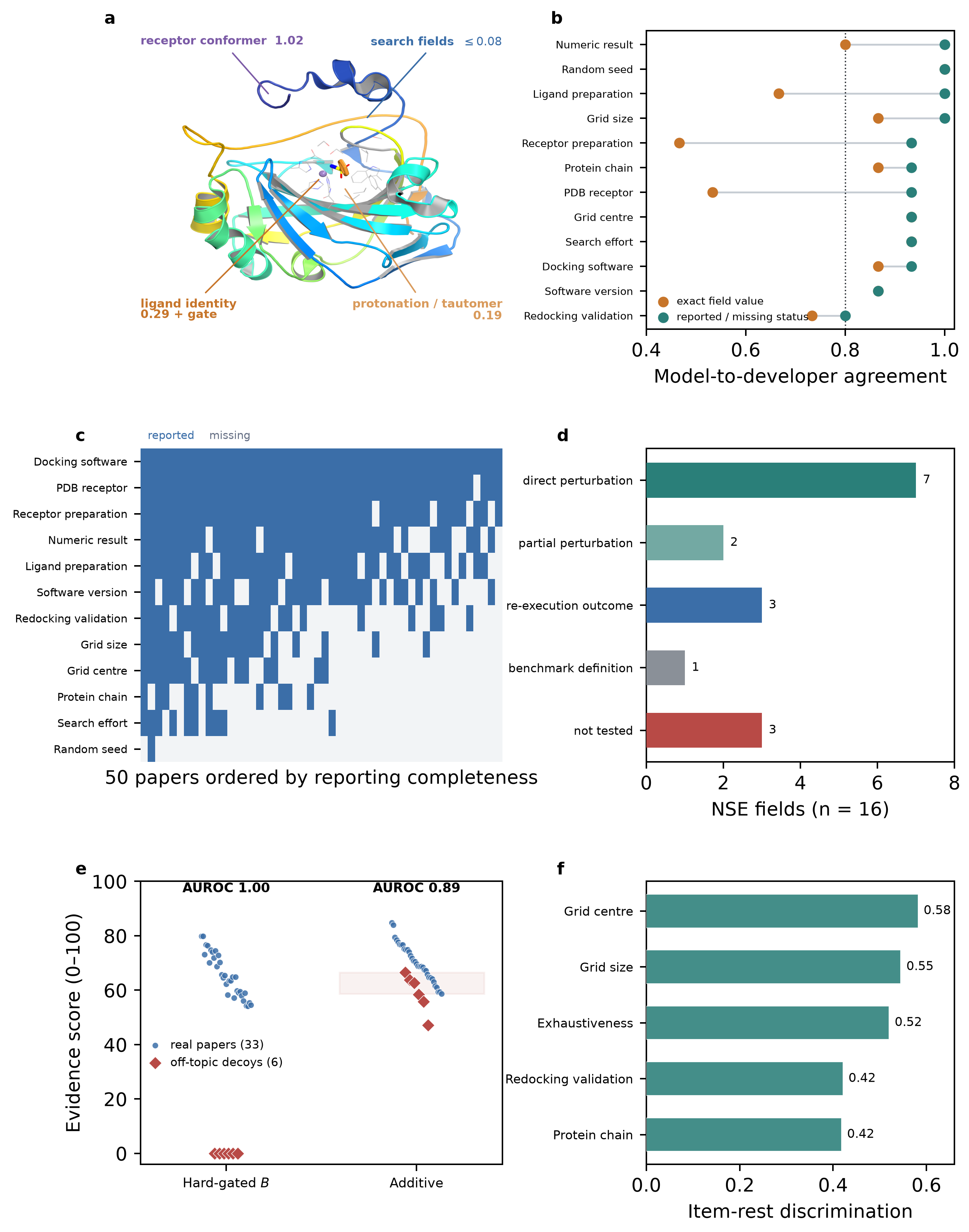}
\caption{Evidence grounding and internal validation of the NSE score.
(a) Representative carbonic anhydrase II structure (PDB 4YX4) with the
FB2 ligand and catalytic zinc, annotated with the largest measured input
effects (Gaspari et al. 2016). (b) Exact-value and
reported-versus-missing agreement between model extraction and final
developer adjudication for 12 fields in 15 papers; the dotted line marks
0.80 agreement. (c) Span-verified 12-field reporting matrix for 50
papers ordered by completeness. (d) Evidence source for all 16 NSE
fields: direct perturbation, partial perturbation, re-execution outcome,
benchmark definition or not tested. (e) Off-topic decoy rejection:
best-evidence scores for 33 on-topic real papers and 6 injected
off-topic decoys under the hard-gated aggregator \(B\) and an additive
aggregator; the hard gate assigns every decoy a zero score (area under
the receiver operating characteristic curve 1.00), whereas the additive
aggregator lets decoys leak into the real-paper range (0.89). (f)
Reliability of the reporting-readiness subscale: item-rest
discriminations of the five-item core subscale, which has Cronbach
\(\alpha\) = 0.73 (full 12-item \(\alpha\) = 0.62; seven fields were
retained as descriptive checklist items outside this selected
subscale).}
\end{figure}

\textbf{Table 4. Empirical reweighting of the 16 NSE method fields and
the evidence supporting each field. The prespecified tier ranges from 1
to 3, with 3 denoting the highest assigned importance. Where a
controlled perturbation was available, effect size is the median
\(|\Delta S|\) in kcal mol\(^{-1}\). Evidence classes comprise direct
ablation (L1), partial ablation (L1p), reproduction-outcome data (L2), a
definitional criterion (L4) and fields not tested in the present study.}

{\def\LTcaptype{none} 
\begin{longtable}[]{@{}
  >{\raggedright\arraybackslash}p{(\linewidth - 8\tabcolsep) * \real{0.2000}}
  >{\raggedright\arraybackslash}p{(\linewidth - 8\tabcolsep) * \real{0.2000}}
  >{\raggedright\arraybackslash}p{(\linewidth - 8\tabcolsep) * \real{0.2000}}
  >{\raggedright\arraybackslash}p{(\linewidth - 8\tabcolsep) * \real{0.2000}}
  >{\raggedright\arraybackslash}p{(\linewidth - 8\tabcolsep) * \real{0.2000}}@{}}
\toprule\noalign{}
\begin{minipage}[b]{\linewidth}\raggedright
Field
\end{minipage} & \begin{minipage}[b]{\linewidth}\raggedright
Prespecified tier
\end{minipage} & \begin{minipage}[b]{\linewidth}\raggedright
Measured effect (kcal)
\end{minipage} & \begin{minipage}[b]{\linewidth}\raggedright
Evidence source
\end{minipage} & \begin{minipage}[b]{\linewidth}\raggedright
Empirical weight or role
\end{minipage} \\
\midrule\noalign{}
\endhead
\bottomrule\noalign{}
\endlastfoot
PDB receptor & 3.0 & 1.02 & Direct perturbation & 1.02 \\
Protein chain & 2.0 & 1.02 & Partial perturbation & 1.02
(receptor-linked) \\
Ligand identifier & 3.0 & 0.29 & Direct perturbation & prerequisite
(22\% unresolved/ambiguous) \\
Ligand preparation & 2.0 & 0.19 & Partial perturbation & 0.19
(protonation component) \\
Protonation / tautomer & 1.0 & 0.19 & Direct perturbation & 0.19 \\
Grid centre & 3.0 & 0.08 & Direct perturbation & 0.08 \\
Grid size & 3.0 & 0.05 & Direct perturbation & 0.05 \\
Random seed & 1.0 & 0.04 & Direct perturbation & 0.04 (self-consistency
floor) \\
Search effort (exhaustiveness) & 1.0 & 0.01 & Direct perturbation &
0.01 \\
Docking software & 3.0 & not estimated & Benchmark definition & fixed
engine \\
Numeric result & 1.0 & not estimated & Reproduction outcome & outcome
measure (29/37 within tolerance) \\
Redocking validation & 2.0 & not estimated & Reproduction outcome &
reported practice \\
Software version & 2.0 & not estimated & Reproduction outcome &
cross-environment floor, 0.43 \\
Receptor preparation & 2.0 & not estimated & Not tested & requires
direct perturbation \\
Water and ion handling & 1.0 & not estimated & Not tested & requires
direct perturbation \\
Code artifacts & 1.0 & not estimated & Not tested & requires a dedicated
comparison \\
\end{longtable}
}

\subsection{Reproduction outcomes differed descriptively by
executability
stratum}\label{reproduction-outcomes-differed-descriptively-by-executability-stratum}

Among the 37 quality-control-passed re-executions, 29 (78\%;
paper-clustered bootstrap 95\% CI, 63--94\%) were within the 2.0 kcal
mol\(^{-1}\) tolerance (Table 5, Fig. 6). All 6 fully reported E3 claims
were within tolerance, with a median absolute difference of 0.24 kcal
mol\(^{-1}\), which is below the conservative 0.43 kcal mol\(^{-1}\)
cross-environment comparison reference. Among 31 assumption-dependent E2
claims, 23 were within tolerance and the median difference was 0.59 kcal
mol\(^{-1}\). The E3-E2 comparison was not statistically significant; it
was underpowered and confounded by box reporting (Mann-Whitney
\(P=0.138\), \(n_{E3}=6\)). Omitting the box forces the re-executor to
assume it, so E2 mixes the effect of an assumed box with incomplete
reporting. The powered design in Section 3.8 is intended to separate
these effects. Continuous NSE quality was also unrelated to absolute
error (Pearson \(r=0.10\)) or the binary within-tolerance outcome
(point-biserial \(r=-0.06\)). E-class therefore describes whether
re-execution is directly supported by the report; these data do not
establish it as a reproduction predictor.

\begin{figure}
\centering
\includegraphics[width=1\linewidth,height=\textheight,keepaspectratio,alt={Reproduction outcomes and the limits of executability strata. (a) Proportions within 2.0 kcal mol\^{}\{-1\} by E-class with Wilson 95\% confidence intervals. (b) Absolute reported-rerun differences; horizontal segments show medians and dotted and dashed lines denote the 0.04 and 0.43 kcal mol\^{}\{-1\} computational floors. (c) Absolute error versus continuous NSE reporting quality. (d) Paper-level within-tolerance proportions with Wilson intervals; the dashed line marks the overall 78\% proportion. The E3-E2 comparison was underpowered and confounded by box reporting (Mann-Whitney P = 0.138).}]{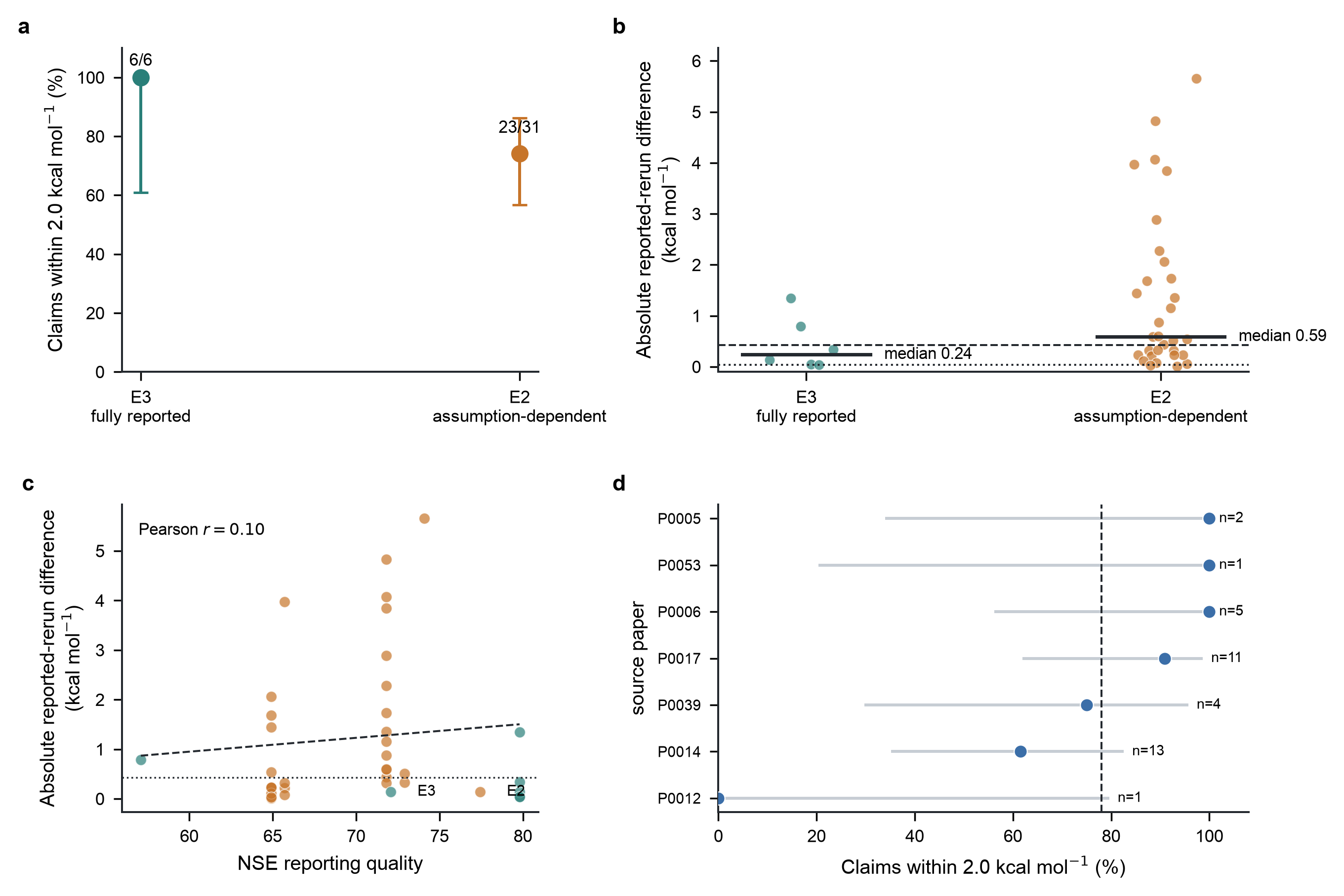}
\caption{Reproduction outcomes and the limits of executability strata.
(a) Proportions within 2.0 kcal mol\(^{-1}\) by E-class with Wilson 95\%
confidence intervals. (b) Absolute reported-rerun differences;
horizontal segments show medians and dotted and dashed lines denote the
0.04 and 0.43 kcal mol\(^{-1}\) computational floors. (c) Absolute error
versus continuous NSE reporting quality. (d) Paper-level
within-tolerance proportions with Wilson intervals; the dashed line
marks the overall 78\% proportion. The E3-E2 comparison was underpowered
and confounded by box reporting (Mann-Whitney \(P = 0.138\)).}
\end{figure}

\FloatBarrier

\textbf{Table 5. Reproduction outcomes by executability stratum among 37
quality-control-passed AutoDock Vina re-executions from 7 papers. E3
denotes a fully reported calculation and E2 denotes a calculation
completed using explicit assumptions. The E3-E2 comparison was
underpowered and confounded by box reporting (Mann-Whitney
\(P = 0.138\), \(n_{E3} = 6\)); the strata are therefore descriptive and
are not interpreted as a reproduction model.}

{\def\LTcaptype{none} 
\begin{longtable}[]{@{}
  >{\raggedright\arraybackslash}p{(\linewidth - 8\tabcolsep) * \real{0.2000}}
  >{\raggedright\arraybackslash}p{(\linewidth - 8\tabcolsep) * \real{0.2000}}
  >{\raggedright\arraybackslash}p{(\linewidth - 8\tabcolsep) * \real{0.2000}}
  >{\raggedright\arraybackslash}p{(\linewidth - 8\tabcolsep) * \real{0.2000}}
  >{\raggedright\arraybackslash}p{(\linewidth - 8\tabcolsep) * \real{0.2000}}@{}}
\toprule\noalign{}
\begin{minipage}[b]{\linewidth}\raggedright
Stratum
\end{minipage} & \begin{minipage}[b]{\linewidth}\raggedright
n
\end{minipage} & \begin{minipage}[b]{\linewidth}\raggedright
Within-2.0
\end{minipage} & \begin{minipage}[b]{\linewidth}\raggedright
Median \textbar reported-rerun\textbar{}
\end{minipage} & \begin{minipage}[b]{\linewidth}\raggedright
95\% CI
\end{minipage} \\
\midrule\noalign{}
\endhead
\bottomrule\noalign{}
\endlastfoot
ALL (QC-passed) & 37 & 29/37 (78\%) & 0.54 & paper-clustered 63--94\% \\
E3 (fully reported) & 6 & 6/6 (100\%) & 0.24 & 0.04--1.07 \\
E2 (box guessed) & 31 & 23/31 (74\%) & 0.59 & 0.32--1.44 \\
E3 vs E2 test & 6 vs 31 & n/a & Mann-Whitney \(P = 0.138\) &
underpowered; box-confounded \\
By target: 6LU7 (Mpro) & 35 & 27/35 & 0.59 & not estimated \\
By target: 6Y2E (Mpro) & 2 & 2/2 & 0.42 & not estimated \\
\end{longtable}
}

\subsection{Reproduction on independent
infrastructure}\label{reproduction-on-independent-infrastructure}

To test whether the reproduction outcomes depended on the local
computing environment, we re-executed the reproduction harness on an
independent cloud virtual machine (AutoDock Vina 1.2.7, seed 200,
exhaustiveness 16), resolving ligand structures and re-preparing
receptors from scratch. Of 37 quality-control-passed re-executions, 27
(73\%) were within the 2.0 kcal mol\(^{-1}\) tolerance (Fig. 7a),
consistent with the 78\% obtained locally, and the fully reported E3
claims again reproduced more often (4 of 5) than the
assumption-dependent E2 claims (23 of 32). For the 35 claims docked on
both machines, the virtual-machine and local re-run affinities agreed
closely, with a median absolute difference of 0.18 kcal mol\(^{-1}\) and
all 35 within 1.0 kcal mol\(^{-1}\) (Fig. 7b); the within-tolerance
verdict flipped for a single claim (3\%). A separate, more divergent
environment pair had previously given a larger cross-environment
difference (0.43 kcal mol\(^{-1}\), n = 15), so cross-environment
variance is environment-pair-dependent and spans roughly 0.18 to 0.43
kcal mol\(^{-1}\), with the larger value serving as a conservative
score-variation reference (Section 3.3). The reproduction rate is
therefore not an artifact of a single machine, whereas the residual
cross-environment variance, which exceeds the 0.04 kcal mol\(^{-1}\)
within-machine self-consistency floor, is the quantity a multi-site
benchmark must report.

\begin{figure}
\centering
\includegraphics[width=1\linewidth,height=\textheight,keepaspectratio,alt={Reproduction on independent infrastructure (cloud virtual machine, AutoDock Vina 1.2.7, seed 200, exhaustiveness 16). (a) Reported versus virtual-machine re-run affinity for 37 quality-control-passed claims; filled points fall within the 2.0 kcal mol\^{}\{-1\} tolerance band and open points fall outside (27 of 37, 73\%). (b) Local versus virtual-machine re-run affinity for the 35 claims docked on both machines. (c) Bland-Altman comparison of virtual-machine and local affinities; solid and dashed lines show the mean bias and 95\% limits of agreement. (d) Concordance of local and virtual-machine within-tolerance verdicts (34 of 35).}]{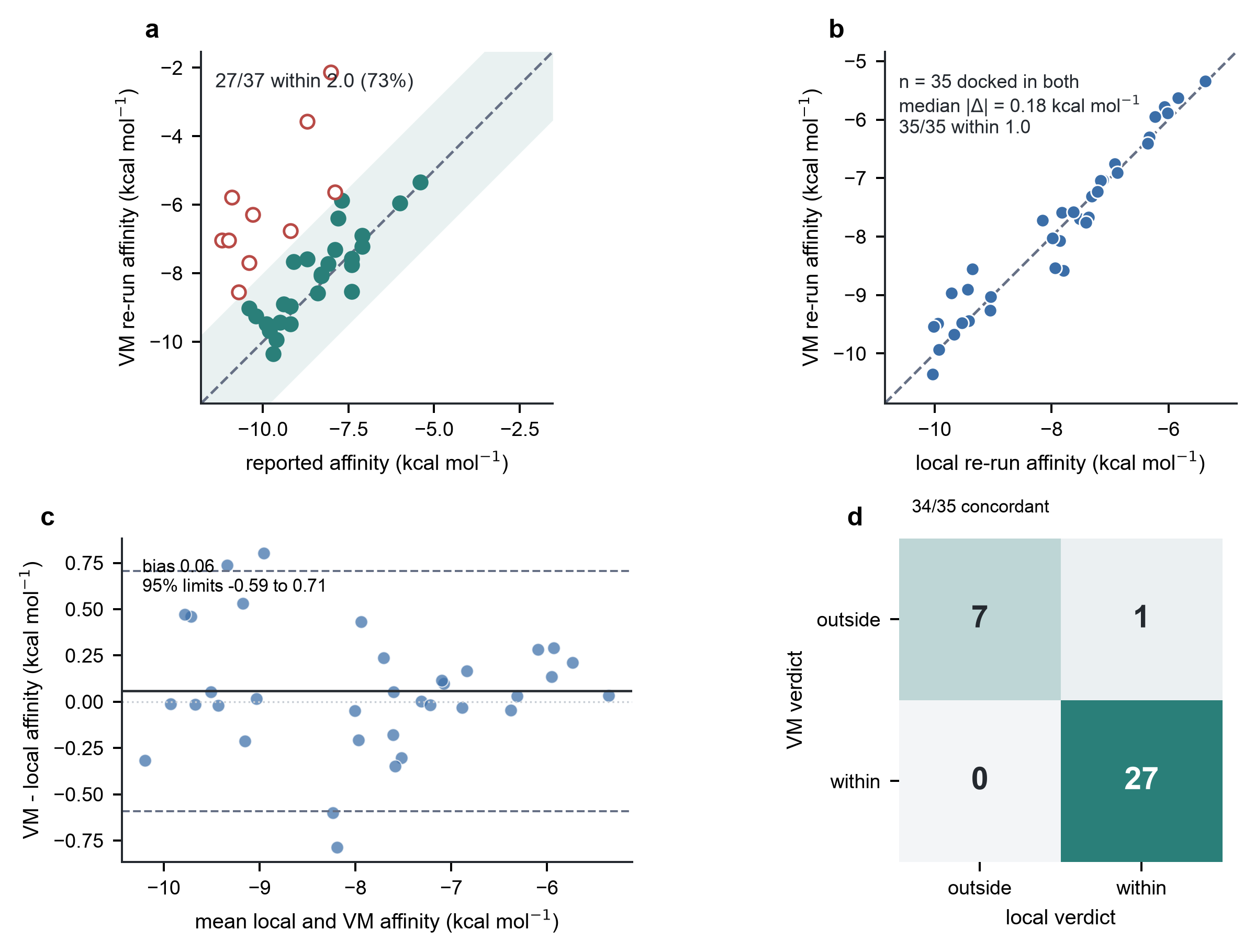}
\caption{Reproduction on independent infrastructure (cloud virtual
machine, AutoDock Vina 1.2.7, seed 200, exhaustiveness 16). (a) Reported
versus virtual-machine re-run affinity for 37 quality-control-passed
claims; filled points fall within the 2.0 kcal mol\(^{-1}\) tolerance
band and open points fall outside (27 of 37, 73\%). (b) Local versus
virtual-machine re-run affinity for the 35 claims docked on both
machines. (c) Bland-Altman comparison of virtual-machine and local
affinities; solid and dashed lines show the mean bias and 95\% limits of
agreement. (d) Concordance of local and virtual-machine within-tolerance
verdicts (34 of 35).}
\end{figure}

\section{Discussion}\label{discussion}

This study separates reporting frequency from the measured influence of
individual docking inputs. Search-space fields were among the least
frequently reported, but their perturbation effects were small under the
tested Vina protocol. Receptor structure produced the largest graded
effect, and ligand identity introduced both score variation and a
categorical resolvability constraint. The receptor-versus-search
contrast was retained in the Vinardo check and the 12-target extension;
ligand-identity and protonation effects were not retested across all
these settings. These findings extend previous evidence that docking
outcomes vary across receptor structures and programs (Warren et al.
2006; Wang et al. 2016; Francoeur et al. 2020) by quantifying
input-specific effects for a reporting and re-execution framework. The
reported targets underlying these analyses span diverse protein folds
and cognate-ligand chemotypes, from carbonic anhydrase and serine
proteases to streptavidin, aldose reductase, cathepsin K and a BRD4
bromodomain (Figure 8), establishing the structural breadth over which
reporting completeness and re-execution were evaluated.

\begin{figure}
\centering
\includegraphics[width=1\linewidth,height=\textheight,keepaspectratio,alt={Structural scope of the benchmark. (a) A representative re-execution target, cathepsin K (PDB 4X6H) bound to its cognate nitrile inhibitor, drawn as the grey receptor cartoon, the translucent binding-pocket surface, the steel-blue search-volume box centred on the cognate ligand and the element-coloured ligand sticks; the four annotated elements are the inputs a docking calculation must specify, and the 2D structure shows the cognate ligand. (b) Eight reported benchmark targets spanning diverse protein folds and cognate-ligand chemotypes: carbonic anhydrase II (PDB 4WW6, 4PZH), alpha thrombin (2BVR), cationic trypsin (5MNN, 2AYW), streptavidin (2F01), aldose reductase (2PZN) and a BRD4 bromodomain (4QB3), each with its cognate ligand shown as a 2D structure. All panels show reported cross-docking or re-execution targets rendered from deposited coordinates; the search-volume boxes are illustrative volumes centred on the cognate ligand, as in Figure 1, and are not new docking results.}]{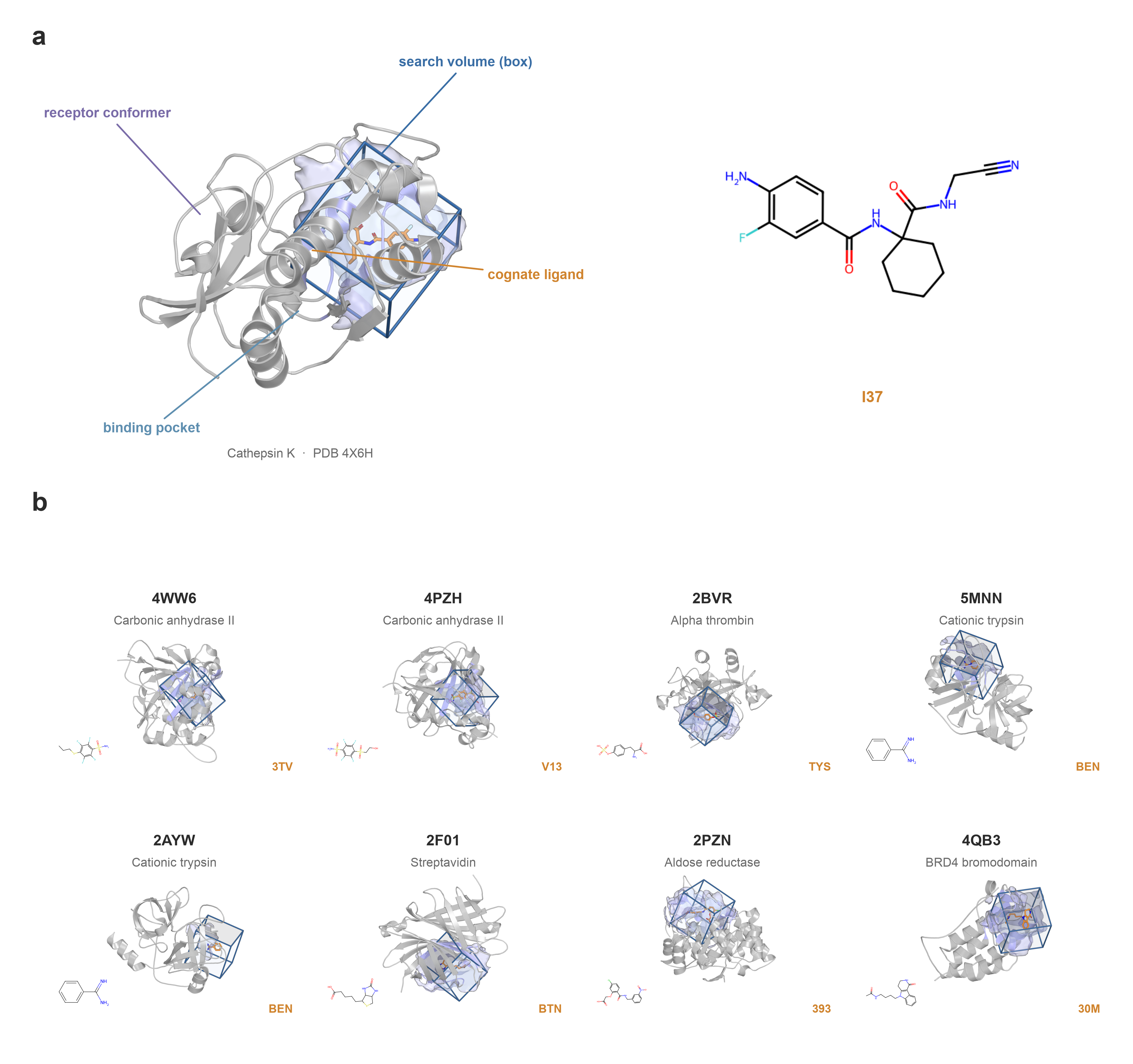}
\caption{Structural scope of the benchmark. (a) A representative
re-execution target, cathepsin K (PDB 4X6H) bound to its cognate nitrile
inhibitor, drawn as the grey receptor cartoon, the translucent
binding-pocket surface, the steel-blue search-volume box centred on the
cognate ligand and the element-coloured ligand sticks; the four
annotated elements are the inputs a docking calculation must specify,
and the 2D structure shows the cognate ligand. (b) Eight reported
benchmark targets spanning diverse protein folds and cognate-ligand
chemotypes: carbonic anhydrase II (PDB 4WW6, 4PZH), alpha thrombin
(2BVR), cationic trypsin (5MNN, 2AYW), streptavidin (2F01), aldose
reductase (2PZN) and a BRD4 bromodomain (4QB3), each with its cognate
ligand shown as a 2D structure. All panels show reported cross-docking
or re-execution targets rendered from deposited coordinates; the
search-volume boxes are illustrative volumes centred on the cognate
ligand, as in Figure 1, and are not new docking results.}
\end{figure}

The empirical weights differ from the prespecified tiers. In particular,
grid centre and grid size had smaller measured effects than protonation
(Brink and Exner 2009), whereas receptor structure and ligand identity
received the largest empirical priority. This result complements
guidance that treats docking as conditional modelling dependent on
structural provenance and ligand-state definition (Kittelson et al.
2026). It also addresses a different question from CASF, PDBbind and
DUD-E, which evaluate scoring, ranking or screening performance against
reference data (Su et al. 2019; Wang et al. 2004; Mysinger et al. 2012).
For reporting practice, the present data support prioritising an exact
receptor structure, a resolvable ligand identifier such as a PubChem CID
or InChIKey, and the ligand protonation state.

The main result is an inversion between reporting emphasis and measured
score sensitivity. Grid centre and grid size were reported in 34\% and
42\% of papers yet produced median absolute score changes of no more
than 0.08 kcal mol\(^{-1}\). By contrast, ligand identity changed the
score by 0.29 kcal mol\(^{-1}\) and left 22\% of unique reported ligand
strings outside the automated re-execution gate when the deterministic
resolver found no unique structure. A checklist weighted by reporting
frequency or untested assumptions therefore misallocates attention.
Weighting fields by measured perturbation effects instead prioritises
the exact receptor structure, a machine-resolvable ligand identifier and
the protonation state; search-box parameters remain necessary for
completeness but receive lower empirical priority.

These weights are a measurement rather than a policy. NSE is designed to
return a per-field recommendation about which reporting elements most
affect re-executability, together with the executability class that
indicates whether a claim can be re-run at all; it is not a
certification that a study is correct, nor a calibrated probability that
a study reproduces. The empirical method-field weights summarize score
shifts, whereas the psychometric weights summarize item associations;
neither is a prospectively validated reproduction predictor; keeping the
retained weighting separate from these measurements prevents a result
obtained on 50 audited papers and a few hundred perturbations from being
read as a validated predictor.

The evidence supports three levels of inference. First, the chosen hard
gate deterministically excludes below-threshold records, and a
decoy-injection test checks that implementation: the hard gate assigns a
zero score to every injected off-topic record (area under the receiver
operating characteristic curve of 1.00), whereas an additive aggregator
lets decoys leak into the real-paper range (0.89) (Figure 5e). Second,
the 50-paper audit supports the reliability-weighted readiness scale,
its internal consistency (Cronbach \(\alpha=0.73\) on the five-item core
subscale; Figure 5f) and its stability to the weighting choice. Third,
the controlled perturbations support the determinant hierarchy across
two scoring functions and 12 targets. The 78\% within-tolerance
reproduction rate is observed but descriptive. The present data do not
support an item-response model with per-paper intervals, a powered
contrast between executability strata or a calibrated probability of
reproduction. Cross-environment variation affects outcome stability near
the energy cutoff, but does not create a resolution limit in probability
units. We therefore separate established measurements from analyses
deferred to the larger campaign specified in Section 3.8.

The evidence-class assignment is necessary because the 16 NSE fields
were not evaluated by a single design. Direct perturbations support 7
fields, partial perturbations support 2, reproduction outcomes describe
3, and the benchmark definition determines 1. Three fields remain
untested. Keeping these sources separate prevents internal consistency
of the reporting scale (Cronbach 1951; Tavakol and Dennick 2011) from
being interpreted as evidence that the scale predicts reproduction.
Similarly, the 78\% within-tolerance proportion is a descriptive result
and not a calibrated probability.

The conclusions apply to computational re-execution of box-based docking
calculations (Peng 2011; Stodden et al. 2016). They do not address
wet-lab replication. The generalization test compared Vina and Vinardo
across 12 targets. Because these scoring functions share the same
box-based search engine and differ only in scoring terms, this analysis
isolates scoring within one search paradigm. Other docking
implementations such as smina (Koes et al. 2013) and the AutoDock4
Lamarckian genetic algorithm (Morris et al. 2009) remain untested. We
also did not evaluate boxless pose prediction with DiffDock or
CNN-assisted scoring with gnina (Corso et al. 2023; McNutt et al. 2021).
The reproduction analysis included only 37 claims from 7 papers, and the
E3 group contained 6 claims; executability-stratum comparisons were
therefore underpowered and confounded by box reporting. Human agreement
on E-class was moderate (\(\kappa=0.47\)), the 3 untested fields require
dedicated perturbations, and the audit completeness estimates describe
the sampled corpus. The verification-priority score was not validated
against expert labels.

Future evaluation should include a larger set of E3 claims in which box
reporting is not confounded with executability, direct perturbations of
the 3 untested fields, boxless methods such as DiffDock, and
CNN-assisted docking methods such as gnina (Corso et al. 2023; McNutt et
al. 2021). Expert pairwise judgements are also required before
interpreting the verification-priority score as a validated ranking
instrument.

\begin{center}\rule{0.5\linewidth}{0.5pt}\end{center}

\section{Data Availability}\label{data-availability}

The arXiv ancillary archive \texttt{reproducibility.zip} contains
retained derived audit tables, annotation forms and agreement summaries,
ligand-resolvability outputs, docking outcome tables, analysis code
available for those retained inputs, figure images and a checksum
ledger. The archive README lists exact input paths, source versions,
executable checks and omissions. The 2026-09-29 package audit recomputes
audit counts, within-tolerance reproduction counts and cross-environment
agreement from retained tables; it does not represent new docking runs
or a fresh full-text collection. The original full build driver and some
intermediate docking tables are unavailable in the live workspace, so
the archive does not claim an executable end-to-end figure or docking
rebuild. Values depending on unavailable intermediates are identified in
the source map and are checked against retained analysis logs.
Third-party article full text is excluded from redistribution. The
public archive omits verbatim article spans; source identifiers and
field-level derived records provide the audit map, and the original
span-verification records remain in local project custody.

Figures 1, 5 and 8 contain original PyMOL structural vignettes of
recorded benchmark complexes: carbonic anhydrase II--FB2 (PDB 4YX4)
(Gaspari et al. 2016) and additional deposited structures (PDB 4X6H,
4WW6, 4PZH, 2BVR, 5MNN, 2AYW, 2F01, 2PZN, 4QB3). These illustrations are
recorded inputs, not additional docking results. Other figure marks use
retained analysis outputs; no generative image model was used. The
original project used the PROV data model (Groth and Moreau 2013) and
RO-Crate (RO-Crate Community, n.d.). This dated release provides its own
source-file checksum ledger rather than presenting the older
project-wide metadata as current.

\section{Declaration of Competing
Interest}\label{declaration-of-competing-interest}

Giap Duc Ha is a scientific cofounder of NewScience, a project
developing verification- and executability-aware tools for computational
scientific claims, and is affiliated with New Science Lab and Nanjing
Medical University. The NSE method evaluated here could inform such
tools. The study object was the reporting and re-execution of docking
claims rather than the NewScience platform, and the derived data, code
and provenance package are supplied so that the findings can be assessed
independently of this commercial interest.

\section{Ethics Statement}\label{ethics-statement}

No human participants, animal experiments or individual-level biomedical
data were used.

\section{Funding}\label{funding}

No external funding was declared for this study.

\section{Acknowledgments}\label{acknowledgments}

A developer-reviewer provided the manual reporting-audit adjudication
recorded in the retained annotation files. The developer-reviewer
retained final label authority; Gemini and GPT outputs were
cross-checks, as described in Methods.

\appendix

\section{Admissible best-evidence
aggregators}\label{admissible-best-evidence-aggregators}

For a separable score \(F(R,Q)=100g(R)h(Q)\) and a nonzero quality
factor, the requirement \(F(R,Q)\to0\) as \(R\to0\) implies \(g(0)=0\)
under continuity at zero. This excludes a positively quality-weighted
additive score. It does not exclude \(100\sqrt{RQ}\), which has
\(g(R)=\sqrt{R}\) and \(h(Q)=\sqrt{Q}\). Whether one missing quality
artifact vetoes a score depends on the construction of \(Q\), not on
this outer aggregator alone.

The selected rule \(F(R,Q)=100\mathbf{1}[R\ge\tau_R]RQ\) is monotone and
separable and enforces exact exclusion below \(\tau_R>0\). Its jump at
the threshold means that it does not satisfy global continuity. Section
2.5 therefore permits this threshold discontinuity explicitly. Other
continuous zero-relevance functions also meet the stated conditions; the
selected hard gate is an operational design choice. Decoy exclusion
checks its implementation and is not an independent test of scientific
validity.

\section{Estimability of the measurement model and the powered
design}\label{estimability-of-the-measurement-model-and-the-powered-design}

The item weights of Section 2.4 are derived from the corrected
item--rest correlation because a two-parameter logistic item-response
model is not stably identifiable at the audit sample size. The
inferences a fitted model would unlock scale with the number of audited
claims \(N\). At \(N\approx50\), the corpus supports Cronbach's alpha,
item--rest weights, the readiness point score and the weight-robustness
check, but not a per-paper standard error, item-fit statistics or tests
for differential item functioning. At a planning target of
\(N\approx250\), a marginal-maximum-likelihood two-parameter logistic
fit could be evaluated for stability; an adequate fit would yield item
discriminations and difficulties, an expected-a-posteriori readiness
estimate with a posterior standard deviation and hence a per-paper
confidence interval, together with infit and outfit statistics and
differential-item-functioning tests across targets and years. A bifactor
extension that separates a general rigour factor from method and
provenance factors requires a larger corpus with at least three items
per factor. Because this scale is estimated from audit reading rather
than from docking, it can be advanced independently of the re-execution
campaign.

The power calculations of Section 3.8 combine three requirements. The
two-proportion contrast between fully reported and assumption-dependent
claims needs about 29 claims per group before clustering and about 40
after the paper-level design effect. The calibration map needs about 10
reproduction events per bin across at least 15 paper clusters, or
roughly 100 to 150 quality-control-passed re-executions. The calibration
requirement binds, so about 150 re-executions over at least 15 papers
and three protein families is a planning target for evaluating the
E3-versus-E2 contrast and calibration together, subject to an adequate
cluster-level design and outcome stability; environment-dependent score
variation would need repeated-environment or threshold-sensitivity
assessment; its energy units cannot be converted directly into a
probability-resolution limit. The item-response upgrade proceeds
separately on about 250 audited claims.

\section{References}\label{references}

\protect\phantomsection\label{refs}


\begin{CSLReferences}{1}{1}
\bibitem[\citeproctext]{ref-ambrosio2023mpro}
{Ambrosio, Francesca Alessandra et al.} 2023. {``Targeting {SARS-CoV-2}
Main Protease: A Successful Story Guided by an in Silico Drug
Repurposing Approach.''} \emph{Journal of Chemical Information and
Modeling}, ahead of print.
\url{https://doi.org/10.1021/acs.jcim.3c00282}.

\bibitem[\citeproctext]{ref-baker2016reproducibility}
Baker, Monya. 2016. {``1,500 Scientists Lift the Lid on
Reproducibility.''} \emph{Nature} 533 (7604): 452--54.
\url{https://doi.org/10.1038/533452a}.

\bibitem[\citeproctext]{ref-brazma2001miame}
{Brazma, Alvis, Pascal Hingamp, John Quackenbush, et al.} 2001.
{``Minimum Information about a Microarray Experiment ({MIAME})---Toward
Standards for Microarray Data.''} \emph{Nature Genetics} 29 (4):
365--71. \url{https://doi.org/10.1038/ng1201-365}.

\bibitem[\citeproctext]{ref-tenbrink2009protonation}
Brink, Tim ten, and Thomas E. Exner. 2009. {``Influence of Protonation,
Tautomeric, and Stereoisomeric States on Protein--Ligand Docking
Results.''} \emph{Journal of Chemical Information and Modeling} 49 (6):
1535--46. \url{https://doi.org/10.1021/ci800420z}.

\bibitem[\citeproctext]{ref-burley2018rcsb}
Burley, Stephen K., Helen M. Berman, Cole Christie, et al. 2018.
{``{RCSB Protein Data Bank}: Sustaining a Living Digital Data Resource
That Enables Breakthroughs in Scientific Research and Biomedical
Education.''} \emph{Protein Science} 27 (1): 316--30.
\url{https://doi.org/10.1002/pro.3331}.

\bibitem[\citeproctext]{ref-corso2023diffdock}
Corso, Gabriele, Hannes Stärk, Bowen Jing, Regina Barzilay, and Tommi
Jaakkola. 2023. \emph{{DiffDock}: Diffusion Steps, Twists, and Turns for
Molecular Docking}. \url{https://doi.org/10.48550/arXiv.2210.01776}.

\bibitem[\citeproctext]{ref-cronbach1951alpha}
Cronbach, Lee J. 1951. {``Coefficient Alpha and the Internal Structure
of Tests.''} \emph{Psychometrika} 16 (3): 297--334.
\url{https://doi.org/10.1007/BF02310555}.

\bibitem[\citeproctext]{ref-dagdelen2024structured}
Dagdelen, John, Alexander Dunn, Sanghoon Lee, et al. 2024. {``Structured
Information Extraction from Scientific Text with Large Language
Models.''} \emph{Nature Communications} 15 (1): 1418.
\url{https://doi.org/10.1038/s41467-024-45563-x}.

\bibitem[\citeproctext]{ref-eberhardt2021vina12}
Eberhardt, Jerome, Diogo Santos-Martins, Andreas F. Tillack, and Stefano
Forli. 2021. {``{AutoDock Vina} 1.2.0: New Docking Methods, Expanded
Force Field, and {Python} Bindings.''} \emph{Journal of Chemical
Information and Modeling} 61 (8): 3891--98.
\url{https://doi.org/10.1021/acs.jcim.1c00203}.

\bibitem[\citeproctext]{ref-flachsenberg2024redocking}
Flachsenberg, Florian, Christiane Ehrt, Torben Gutermuth, and Matthias
Rarey. 2024. {``Redocking the {PDB}.''} \emph{Journal of Chemical
Information and Modeling} 64 (1): 219--37.
\url{https://doi.org/10.1021/acs.jcim.3c01573}.

\bibitem[\citeproctext]{ref-francoeur2020crossdocked}
Francoeur, Paul G., Tomohide Masuda, Jocelyn Sunseri, et al. 2020.
{``Three-Dimensional Convolutional Neural Networks and a Cross-Docked
Data Set for Structure-Based Drug Design.''} \emph{Journal of Chemical
Information and Modeling} 60 (9): 4200--4215.
\url{https://doi.org/10.1021/acs.jcim.0c00411}.

\bibitem[\citeproctext]{ref-gartlehner2024llmextraction}
{Gartlehner, Gerald et al.} 2024. {``Data Extraction for Evidence
Synthesis Using a Large Language Model: A Proof-of-Concept Study.''}
\emph{Research Synthesis Methods} 15 (4): 576--89.
\url{https://doi.org/10.1002/jrsm.1710}.

\bibitem[\citeproctext]{ref-gaspari2016hca2}
Gaspari, Roberto, Christian Rechlin, Andreas Heine, et al. 2016.
{``Kinetic and Structural Insights into the Mechanism of Binding of
Sulfonamides to Human Carbonic Anhydrase by Computational and
Experimental Studies.''} \emph{Journal of Medicinal Chemistry} 59 (9):
4245--56. \url{https://doi.org/10.1021/acs.jmedchem.5b01643}.

\bibitem[\citeproctext]{ref-gebru2021datasheets}
Gebru, Timnit, Jamie Morgenstern, Briana Vecchione, et al. 2021.
{``Datasheets for Datasets.''} \emph{Communications of the ACM} 64 (12):
86--92. \url{https://doi.org/10.1145/3458723}.

\bibitem[\citeproctext]{ref-groth2013prov}
Groth, Paul, and Luc Moreau. 2013. \emph{{PROV-Overview}: An Overview of
the {PROV} Family of Documents}. W3C Working Group Note. W3C.
\url{https://www.w3.org/TR/prov-overview/}.

\bibitem[\citeproctext]{ref-guedes2014docking}
Guedes, Isabella A., Camila S. de Magalhães, and Laurent E. Dardenne.
2014. {``Receptor--Ligand Molecular Docking.''} \emph{Biophysical
Reviews} 6 (1): 75--87. \url{https://doi.org/10.1007/s12551-013-0130-2}.

\bibitem[\citeproctext]{ref-jain2008bias}
Jain, Ajay N. 2008. {``Bias, Reporting, and Sharing: Computational
Evaluations of Docking Methods.''} \emph{Journal of Computer-Aided
Molecular Design} 22 (3--4): 201--12.
\url{https://doi.org/10.1007/s10822-007-9151-x}.

\bibitem[\citeproctext]{ref-kim2025pubchem}
Kim, Sunghwan, Jie Chen, Tiejun Cheng, et al. 2025. {``{PubChem} 2025
Update.''} \emph{Nucleic Acids Research} 53 (D1): D1516--25.
\url{https://doi.org/10.1093/nar/gkae1059}.

\bibitem[\citeproctext]{ref-kittelson2026reproducibility}
Kittelson, Katiana Simões, Allana C. F. Martins, Raquel Possemozer
Santos, Gizele Celante, and Roberto da Silva Gomes. 2026.
{``Reproducibility, Validation, and Failure Modes Across Classical and
{AI}-Driven Molecular Docking.''} \emph{Journal of Computer-Aided
Molecular Design} 40 (1): 137.
\url{https://doi.org/10.1007/s10822-026-00849-8}.

\bibitem[\citeproctext]{ref-koes2013smina}
Koes, David Ryan, Matthew P. Baumgartner, and Carlos J. Camacho. 2013.
{``Lessons Learned in Empirical Scoring with Smina from the {CSAR} 2011
Benchmarking Exercise.''} \emph{Journal of Chemical Information and
Modeling} 53 (8): 1893--904. \url{https://doi.org/10.1021/ci300604z}.

\bibitem[\citeproctext]{ref-konet2024twollms}
{Konet, Amanda et al.} 2024. {``Performance of Two Large Language Models
for Data Extraction in Evidence Synthesis.''} \emph{Research Synthesis
Methods} 15 (5): 818--24. \url{https://doi.org/10.1002/jrsm.1732}.

\bibitem[\citeproctext]{ref-landrum2025rdkit}
Landrum, Greg, Paolo Tosco, Brian Kelley, Ricardo Rodriguez, David
Cosgrove, and RDKit Contributors. 2026. \emph{{RDKit}: Open-Source
Cheminformatics, Release 2025.09.5}. V. 2025.09.5. Zenodo, released.
\url{https://doi.org/10.5281/zenodo.18428170}.

\bibitem[\citeproctext]{ref-mandour2020mpro}
{Mandour, Yasmine M. et al.} 2022. {``A Multi-Stage Virtual Screening of
{FDA}-Approved Drugs Reveals Potential Inhibitors of {SARS-CoV-2} Main
Protease.''} \emph{Journal of Biomolecular Structure and Dynamics},
ahead of print. \url{https://doi.org/10.1080/07391102.2020.1837680}.

\bibitem[\citeproctext]{ref-mcnutt2021gnina}
McNutt, Andrew T., Paul Francoeur, Rishal Aggarwal, et al. 2021.
{``{GNINA} 1.0: Molecular Docking with Deep Learning.''} \emph{Journal
of Cheminformatics} 13 (1): 43.
\url{https://doi.org/10.1186/s13321-021-00522-2}.

\bibitem[\citeproctext]{ref-mitchell2019modelcards}
Mitchell, Margaret, Simone Wu, Andrew Zaldivar, et al. 2019. {``Model
Cards for Model Reporting.''} \emph{Proceedings of the Conference on
Fairness, Accountability, and Transparency ({FAT*})}, 220--29.
\url{https://doi.org/10.1145/3287560.3287596}.

\bibitem[\citeproctext]{ref-morris2009autodock4}
Morris, Garrett M., Ruth Huey, William Lindstrom, et al. 2009.
{``{AutoDock4} and {AutoDockTools4}: Automated Docking with Selective
Receptor Flexibility.''} \emph{Journal of Computational Chemistry} 30
(16): 2785--91. \url{https://doi.org/10.1002/jcc.21256}.

\bibitem[\citeproctext]{ref-mysinger2012dude}
Mysinger, Michael M., Michael Carchia, John J. Irwin, and Brian K.
Shoichet. 2012. {``Directory of Useful Decoys, Enhanced ({DUD-E}):
Better Ligands and Decoys for Better Benchmarking.''} \emph{Journal of
Medicinal Chemistry} 55 (14): 6582--94.
\url{https://doi.org/10.1021/jm300687e}.

\bibitem[\citeproctext]{ref-page2021prisma}
{Page, Matthew J. et al.} 2021. {``The {PRISMA} 2020 Statement: An
Updated Guideline for Reporting Systematic Reviews.''} \emph{BMJ} 372:
n71. \url{https://doi.org/10.1136/bmj.n71}.

\bibitem[\citeproctext]{ref-peng2011reproducible}
Peng, Roger D. 2011. {``Reproducible Research in Computational
Science.''} \emph{Science} 334 (6060): 1226--27.
\url{https://doi.org/10.1126/science.1213847}.

\bibitem[\citeproctext]{ref-peraltamoreno2023mpro}
{Peralta-Moreno, María N. et al.} 2023. {``Autochthonous {Peruvian}
Natural Plants as Potential {SARS-CoV-2} {Mpro} Main Protease
Inhibitors.''} \emph{Pharmaceuticals} 16: 585.
\url{https://doi.org/10.3390/ph16040585}.

\bibitem[\citeproctext]{ref-perciedusert2020arrive}
{Percie du Sert, Nathalie, Viki Hurst, Amrita Ahluwalia, et al.} 2020.
{``The {ARRIVE} Guidelines 2.0: Updated Guidelines for Reporting Animal
Research.''} \emph{PLOS Biology} 18 (7): e3000410.
\url{https://doi.org/10.1371/journal.pbio.3000410}.

\bibitem[\citeproctext]{ref-quiroga2016vinardo}
Quiroga, Rodrigo, and Marcos A. Villarreal. 2016. {``Vinardo: A Scoring
Function Based on {AutoDock Vina} Improves Scoring, Docking, and Virtual
Screening.''} \emph{PLOS ONE} 11 (5): e0155183.
\url{https://doi.org/10.1371/journal.pone.0155183}.

\bibitem[\citeproctext]{ref-rocrate}
RO-Crate Community. n.d. \emph{{RO-Crate} 1.1 Specification}.
\url{https://www.researchobject.org/ro-crate/}.

\bibitem[\citeproctext]{ref-rule2019tensimplerules}
{Rule, Adam et al.} 2019. {``Ten Simple Rules for Writing and Sharing
Computational Analyses in {Jupyter} Notebooks.''} \emph{PLOS
Computational Biology}, ahead of print.
\url{https://doi.org/10.1371/journal.pcbi.1007007}.

\bibitem[\citeproctext]{ref-samuel2023reproducibility}
Samuel, Sheeba, and Daniel Mietchen. 2023. \emph{Computational
Reproducibility of {Jupyter} Notebooks from Biomedical Publications}.
\url{https://arxiv.org/abs/2308.07333}.

\bibitem[\citeproctext]{ref-sandve2013tensimplerules}
Sandve, Geir Kjetil, Anton Nekrutenko, James Taylor, and Eivind Hovig.
2013. {``Ten Simple Rules for Reproducible Computational Research.''}
\emph{PLoS Computational Biology} 9 (10): e1003285.
\url{https://doi.org/10.1371/journal.pcbi.1003285}.

\bibitem[\citeproctext]{ref-santosmartins2025meeko}
Santos-Martins, Diogo, Yiran He, Jerome Eberhardt, et al. 2025.
{``Meeko: Molecule Parametrization and Software Interoperability for
Docking and Beyond.''} \emph{Journal of Chemical Information and
Modeling} 65 (24): 13045--50.
\url{https://doi.org/10.1021/acs.jcim.5c02271}.

\bibitem[\citeproctext]{ref-sisakht2021mpro}
{Sisakht, Mohsen et al.} 2021. {``Plant-Derived Chemicals as Potential
Inhibitors of {SARS-CoV-2} Main Protease (6LU7), a Virtual Screening
Study.''} \emph{Phytotherapy Research}, ahead of print.
\url{https://doi.org/10.1002/ptr.7041}.

\bibitem[\citeproctext]{ref-stodden2016enhancing}
Stodden, Victoria, Marcia McNutt, David H. Bailey, et al. 2016.
{``Enhancing Reproducibility for Computational Methods.''}
\emph{Science} 354 (6317): 1240--41.
\url{https://doi.org/10.1126/science.aah6168}.

\bibitem[\citeproctext]{ref-su2019casf}
Su, Minyi, Qifan Yang, Yu Du, et al. 2019. {``Comparative Assessment of
Scoring Functions: The {CASF-2016} Update.''} \emph{Journal of Chemical
Information and Modeling} 59 (2): 895--913.
\url{https://doi.org/10.1021/acs.jcim.8b00545}.

\bibitem[\citeproctext]{ref-tavakol2011cronbach}
Tavakol, Mohsen, and Reg Dennick. 2011. {``Making Sense of {Cronbach's}
Alpha.''} \emph{International Journal of Medical Education} 2: 53--55.
\url{https://doi.org/10.5116/ijme.4dfb.8dfd}.

\bibitem[\citeproctext]{ref-trott2010vina}
Trott, Oleg, and Arthur J. Olson. 2010. {``{AutoDock Vina}: Improving
the Speed and Accuracy of Docking with a New Scoring Function, Efficient
Optimization, and Multithreading.''} \emph{Journal of Computational
Chemistry} 31 (2): 455--61. \url{https://doi.org/10.1002/jcc.21334}.

\bibitem[\citeproctext]{ref-wang2004pdbbind}
Wang, Renxiao, Xueliang Fang, Yipin Lu, and Shaomeng Wang. 2004. {``The
{PDBbind} Database: Collection of Binding Affinities for Protein--Ligand
Complexes with Known Three-Dimensional Structures.''} \emph{Journal of
Medicinal Chemistry} 47 (12): 2977--80.
\url{https://doi.org/10.1021/jm030580l}.

\bibitem[\citeproctext]{ref-wang2016tendocking}
Wang, Zhe, Huiyong Sun, Xiaojun Yao, et al. 2016. {``Comprehensive
Evaluation of Ten Docking Programs on a Diverse Set of Protein--Ligand
Complexes: The Prediction Accuracy of Sampling Power and Scoring
Power.''} \emph{Physical Chemistry Chemical Physics} 18 (18): 12964--75.
\url{https://doi.org/10.1039/c6cp01555g}.

\bibitem[\citeproctext]{ref-warren2006assessment}
Warren, Gregory L., C. Webster Andrews, Anna-Maria Capelli, et al. 2006.
{``A Critical Assessment of Docking Programs and Scoring Functions.''}
\emph{Journal of Medicinal Chemistry} 49 (20): 5912--31.
\url{https://doi.org/10.1021/jm050362n}.

\bibitem[\citeproctext]{ref-wilkinson2016fair}
{Wilkinson, Mark D., Michel Dumontier, IJsbrand Jan Aalbersberg, et al.}
2016. {``The {FAIR} Guiding Principles for Scientific Data Management
and Stewardship.''} \emph{Scientific Data} 3: 160018.
\url{https://doi.org/10.1038/sdata.2016.18}.

\bibitem[\citeproctext]{ref-yisha2026llmrct}
{Yisha, Z. et al.} 2026. {``Assessing Data Extraction in Randomized
Clinical Trials with Large Language Models.''} \emph{BMC Medical
Research Methodology}, ahead of print.
\url{https://doi.org/10.1186/s12874-025-02729-5}.

\bibitem[\citeproctext]{ref-zajacek2024mpro}
{Zajáček, Dávid et al.} 2024. {``Compromise in Docking Power of Liganded
Crystal Structures of {Mpro} {SARS-CoV-2} Surpasses 90\% Success
Rate.''} \emph{Journal of Chemical Information and Modeling} 64 (5):
1628--43. \url{https://doi.org/10.1021/acs.jcim.3c01552}.

\end{CSLReferences}
\end{document}